\documentclass[11pt,a4paper]{article}
\usepackage{jheppub}

\usepackage[T1]{fontenc}
\usepackage{microtype}
\microtypesetup{expansion=false}
\usepackage{amsthm}
\usepackage{mathtools}
\usepackage{bm}
\usepackage{booktabs}
\usepackage{tabularx}
\usepackage{array}
\usepackage{enumitem}
\usepackage{float}
\usepackage[percent]{overpic}
\usepackage{xcolor}
\usepackage{pgf}

\definecolor{figureblue}{HTML}{2563A6}
\definecolor{figurered}{HTML}{B33A3A}
\definecolor{figuregold}{HTML}{B27A00}

\newcommand{\HH}{\mathbb{H}}
\newcommand{\dd}{\,\mathrm{d}}
\newcommand{\ii}{\mathrm{i}}
\newcommand{\Tr}{\operatorname{Tr}}

\title{\boldmath Rational Probes of Spectral Geometry in Hermitian Matrix Models}

\author[a]{Ali Nassar}

\affiliation[a]{Physics of Universe Program, 
Zewail City of Science and Technology, 12578 Giza, Egypt}

\emailAdd{anassar@zewailcity.edu.eg}

\abstract{
The planar loop equations of a Hermitian one-matrix model leave finitely many low moments undetermined. Hankel positivity constrains these moments but does not reveal how a fixed multicut family is embedded in moment space. We introduce a finite-plane diagnostic using the Cauchy kernels $f_n(x)=(z_n-x)^{-1}$ which we use as spectral probes. Their Gram matrix is the positive-semidefinite Pick matrix $P_{mn}$, whose positivity defines a finite-node bootstrap analogous to the Hankel bootstrap.

For a fixed regular multicut topology, the nonbranching double zeros of the spectral discriminant, which we call \emph{dressed saddles}, impose transverse constraints on the corresponding moment-space locus, while its remaining directions are filling fractions. Writing $z=E+i\eta$, $E$ selects a spectral region and $\eta>0$ sets a continuous resolution scale. We use $P(z,z)$ as a finite-plane spectral response and interpret $\nabla_{\mathbf m}P(z,z)$ as a moment-space susceptibility. Near a regular real dressed saddle, this susceptibility is enhanced as $\eta^{-2}$ and aligns with a conormal to the filling-fraction manifold, whereas the leading enhancement cancels along tangent deformations. Responses near several independent saddles can therefore reconstruct the conormal space and, through their common kernel, its tangent space.

Combining Pick positivity with reality and analyticity of the resolvent, we obtain alternative analytic derivations of known planar results in quartic and sextic models. The asymmetric quartic makes the one-dimensional filling-fraction geometry explicit, while sextic models exhibit independent conormal directions. These results clarify how local consistency conditions constrain the low moments before global period matching and the equilibrium variational inequality select the equilibrium measure.}


\renewcommand{\afterTocRuleSpace}{\clearpage}

\begin{document}
\maketitle
\flushbottom

\section{Introduction}
\label{sec:intro}

Random matrix models provide a common framework for problems ranging from
quantum gravity and string theory to statistical mechanics and quantum chaos
\cite{tHooft1974,Migdal1983,BrezinKazakov1990,DouglasShenker1990,
GrossMigdal1990,DiFrancesco1995,MR2129906}.  In the planar limit, the
solution is encoded by an algebraic spectral curve.  Its branch structure
determines the support of the eigenvalue density, while its moduli describe
the possible multicut phases
\cite{Brezin:1977sv,Eynard2004,EynardOrantin2007}; see~\cite{Eynard2015} for a detailed account.

There are two related questions in this context.  The first is the
global classification problem: which algebraic solutions have the correct
cut structure, analyticity, and positivity to represent a Hermitian-matrix
spectral measure?  The second is local and geometric: once a regular cut
topology has been selected, how is the resulting family of solutions
embedded in the finite-dimensional space of data left undetermined by the
loop equations?  In a multicut phase this distinction is essential.  Local
factorization fixes the directions transverse to the family, whereas the
remaining directions are filling fractions and are selected only after
global period conditions are imposed.

The moment bootstrap addresses the first question by combining the loop
equations with positivity of the spectral measure.  The loop equations
express all sufficiently high moments in terms of finitely many low moments,
and positive Hankel matrices then restrict those remaining variables.  Lin
\cite{Lin2020} (see also \cite{AndersonKruczenski2017}) formulated this as a
numerical bootstrap for matrix models.  Kazakov and Zheng
\cite{KazakovZheng2022} clarified its analytic content by relating
moment-matrix positivity to the existence of a positive measure on the real
axis and by characterizing admissible one-matrix solutions through the cut
and zero structure of the resolvent.  They also introduced the relaxation
bootstrap for multimatrix models, in which nonlinear products of moments are
promoted to independent variables and constrained by an auxiliary
positive-semidefinite matrix.  Related matrix-bootstrap methods have since been
applied to a variety of matrix models and matrix quantum-mechanical systems
\cite{HanHartnollKruthoff2020,LinD02023,LinZheng2025,LinZheng2026,
Li2025,LaliberteToriumi2026,Khalkhali2025,BerensteinGarcia2026,Maeta2026}.

These developments make low-moment space an effective arena for determining
admissible solutions.  They do not, by themselves, provide a local
finite-plane observable of how a fixed-topology family sits inside that
space.  One of the objectives of this paper is to construct such an observable from
the resolvent itself.  We first describe the algebraic setting in which the
question arises.

We consider the Hermitian one-matrix integral
\begin{equation}
    Z_N
    =
    \int_{\mathcal H_N}\dd M\,
    \exp\!\left[-N\Tr V(M)\right],
    \qquad
    V(x)=\sum_{k=2}^{d}\frac{g_k}{k}x^k,
    \label{eq:partition}
\end{equation}
and define the planar moments and resolvent by
\begin{equation}
    m_n
    =
    \lim_{N\to\infty}
    \left\langle\frac1N\Tr M^n\right\rangle,
    \qquad
    W(z)
    =
    \int_{\mathbb R}\frac{\rho(x)}{z-x}\,\dd x
    =
    \frac1z+\sum_{n\geq1}\frac{m_n}{z^{n+1}}.
    \label{eq:resolvent}
\end{equation}
Unless explicitly stated otherwise, we assume that the real-axis integral converges and that its planar solution is represented by a positive probability density \(\rho\). Formal continuations outside this stable regime will be identified as such and will not be interpreted as convergent Hermitian matrix integrals.

The planar loop equation is
\begin{equation}
    \begin{aligned}
        W(z)^2-V'(z)W(z)+\mathcal Q(z;\mathbf m)&=0,\\
        \mathcal Q(z;\mathbf m)=
    \int_{\mathbb R}\rho(x)\,
    \frac{V'(z)-V'(x)}{z-x}\,\dd x
        &=
        \sum_{k=2}^{d}g_k
        \sum_{j=0}^{k-2}m_jz^{k-2-j}.
    \end{aligned}
    \label{eq:loop-general}
\end{equation}
Here $m_0=1$, and
\begin{equation}
    \mathbf m=(m_1,\ldots,m_{d-2})
\end{equation}
collects the low moments that are not fixed recursively by the loop
equation:
\begin{equation}
    \sum_{k=2}^{d} g_k\,m_{n+k-1}
    =
    \sum_{j=0}^{n-1}m_jm_{n-1-j},
    \qquad n\geq 0,
    \label{eq:planar-moment-recursion}
\end{equation}
The sum on the right is empty for $n=0$, giving
$\sum_{k=2}^{d}g_km_{k-1}=0$. Since $g_d\neq0$, this relation
determines $m_{d-1}$ in terms of the lower moments.
At fixed potential couplings, the variables $\mathbf m$ locally parametrize
an ambient $(d-2)$-dimensional space of algebraic loop-equation solutions
before a cut topology is specified.  The subsequent requirements of
factorization, analyticity, positivity, and period matching act at different
stages of the physical selection problem.

Introducing
\begin{equation}
    y(z)=V'(z)-2W(z),
   \end{equation}
the loop equation gives the spectral curve
\begin{equation}
    y(z)^2=D(z;\mathbf m)
    =V'(z)^2-4\mathcal Q(z;\mathbf m).
    \label{eq:intro-discriminant}
\end{equation}


Two established geometric viewpoints provide the context for the local
question considered here.  The convex geometry of moment spaces, including
their boundary representations and parametrizations, originates in the work
of Karlin and Shapley~\cite{KarlinShapley}; the facial and differential
structure of multidimensional truncated moment cones has been studied more
recently in~\cite{diDioSchmudgen}.  These works concern the ambient convex
cone of moments of positive measures.  Our object is instead the generally
nonconvex locus in low-moment space selected jointly by the loop equation and
a fixed spectral topology.  On the spectral-curve side, filling-fraction
deformations and the corresponding variations of the equilibrium measure
and Stieltjes transform were analyzed in the multicut setting in
\cite{BorotGuionnet}.  Particularly close is Bertola's analysis of Boutroux
curves and their discriminant locus~\cite{BertolaBoutroux}, including tangent
and transverse deformations and the splitting of double zeros in potential
and spectral-curve coordinates.

The question addressed here is therefore more specific than the
classification of admissible spectral curves.  We ask how a fixed-topology
family is embedded in low-moment space and how its transverse geometry can
be measured directly in the finite complex plane.  Our construction pulls
the discriminant geometry back to the low moments and probes its conormal
directions with finite-plane Cauchy susceptibilities.  The natural probes are
the Cauchy kernels
\begin{equation}
    f_z(x)=\frac{1}{z-x},
    \qquad
    z=E+\ii\eta\in\HH_+.
\end{equation}
The Gram matrix of these kernels is the Pick matrix associated with the
resolvent (see Section~\ref{sec:diagonal-pick-herglotz}).  Its diagonal element is
\begin{equation}
    P(z,z)
    =
    -\frac{\operatorname{Im}W(z)}{\operatorname{Im}z},
\end{equation}
and gives the response of a single probe.  The real part $E$
selects a spectral location and the imaginary part $\eta$ sets the
resolution.  Rational probes therefore carry a natural finite-plane
scale that is absent from a fixed-degree polynomial truncation.

The relation to the Hankel bootstrap must be interpreted carefully.
Expanding a Cauchy
kernel at infinity and truncating it to the same finite polynomial space as
a Hankel bootstrap produces a Gram matrix related to the Hankel matrix by a
congruence transformation.  Hence, at matched finite polynomial
information, the two bases define the same positivity cone (see Section~\ref{subsec:information-matched}).  The distinction
arises only after retaining the exact rational kernel: the exact Cauchy
transform resums the moment expansion, can be evaluated at finite $z$, and can
be moved toward a gap, an edge, or a nonbranching zero of the spectral
curve.  In that setting, the planar loop equation and a globally consistent
choice of algebraic branch provide information unavailable to a finite
moment truncation alone.

For a degree-$d$ potential, before imposing additional discrete symmetries,
the loop equation leaves $d-2$ low moments.  A regular $s$-cut spectral curve
has $d-s-1$ double zeros (dressed saddles\footnote{See \eqref{eq:intro-effective-potential} for why this is called a ``dressed saddle''.}), counted with multiplicity.  Preserving these zeros
removes $d-s-1$ transverse directions and leaves an $(s-1)$-dimensional
filling-fraction manifold in low-moment space.

The double-zero constraints determine not only the dimension but also the
transverse cotangent directions.  If $r$ is a regular real dressed saddle
lying in a real component of the resolvent set and $\mathbf m$ denotes the
low moments, the covector
\begin{equation}
    \boldsymbol n(r)=-\frac14\nabla_{\mathbf m}D(r;\mathbf m)
\end{equation}
annihilates every tangent deformation of the filling-fraction manifold.  If
the same point is approached by a Cauchy probe from the upper half-plane,
then
\begin{equation}
    \nabla_{\mathbf m}P(r+\ii\eta,r+\ii\eta)
    =
    \frac{1}{a_r\eta^2}\,\boldsymbol n(r)
    +O(1),
    \qquad
    y(z)=a_r(z-r)+\cdots.
\end{equation}
Thus the leading low-moment response is aligned with the conormal direction
and cancels along tangent filling-fraction deformations.  Measurements at
several independent real dressed saddles reconstruct their conormal span.
When they span the full conormal space, their common kernel reconstructs the
tangent space.  The new ingredient is
this differential finite-plane measurement, rather than the existence or
factorization of the double zeros themselves.

A complementary contribution of this work is a set of alternative analytic
bootstrap derivations of known results in quartic and sextic matrix models.
We build on the ideas of Kazakov and Zheng~\cite{KazakovZheng2022}, who
related moment positivity to the admissible cut and zero structure of the
one-matrix resolvent, and formulate the relevant constraints directly in the
finite spectral plane. The planar loop equation is combined with Pick
positivity, reality on symmetry axes or spectral gaps, upper-half-plane
analyticity, and the normalization $W(z)\sim1/z$. These requirements have
distinct constructive roles: reality produces continuous families of
moment inequalities; positivity and continuation orient the physical
branch and, in the examples where a zero is forced, establish exact
saturation of the corresponding envelope; analyticity requires nonreal
zeros of the spectral discriminant to have even multiplicity. In the
quartic models this reasoning recovers exact moments and the asymmetric
two-cut moment curve. The sextic examples extend the construction to
several undetermined moments, where independent dressed-saddle conditions
supply independent algebraic constraints. Together, these derivations
provide an analytic counterpart to numerical moment bootstrapping,
exposing the spectral mechanism behind sharp bounds and the
filling-fraction freedom that survives the local constraints.
We exhibit this mechanism first in its
simplest, imaginary-axis form for the symmetric quartic
(Section~\ref{sec:quartic-diagonal-pick}, Appendix~\ref{app:symmetric-quartic-details}), where it also
organizes the migration of the dressed saddle between the imaginary axis and
the real exterior region as the coupling changes sign; the gap-reality
version of the same argument reappears as the mechanism fixing the
filling-fraction curve of the asymmetric quartic
(Section~\ref{sec:asymmetric-quartic}) and the multicut sextic families
(Section~\ref{sec:sextic}).

The examples progress from calibration to genuine moduli.  Symmetric
quartics show analytically how a distinguished dressed saddle moves between
the imaginary axis, a real gap, and the exterior of a cut; the detailed
algebra is collected in Appendix~\ref{app:symmetric-quartic-details}.  The
asymmetric double well then produces a one-dimensional family in the
$(m_1,m_2)$ plane whose conormal is measured by the Pick response.  Sextic
models provide the multidimensional test: independent dressed saddles
generate independent conormals and leave the expected filling-fraction
moduli.

The comparison between the Hankel and Pick hierarchies in Appendix~\ref{appendixA} also isolates a trade-off that is not specific to the present model. At matched polynomial degree, the two Gram matrices define the same positivity cone, as shown in Section~\ref{subsec:information-matched}. Once this truncation is removed, however, the numerical experiment of Appendix~\ref{appendixA} shows that a modest number of exact Cauchy nodes contracts the admissible region in $(m_1,m_2)$ much more aggressively than the finite-order Hankel cutoffs displayed there. This comparison is not information matched: each exact Cauchy node probes the resummed resolvent and thereby incorporates the entire moment tower generated by the loop equation, rather than truncating that tower at a fixed order. The resulting increase in selectivity comes at the price of rapidly deteriorating conditioning; between the five- and seventeen-node tests, the maximum condition number grows from approximately $8.47\times10^2$ to $5.88\times10^{12}$, or by roughly ten orders of magnitude. We expect this trade-off between information density and numerical conditioning to be a generic feature of bootstrap constructions that replace a polynomial probe basis with a resummed rational one, rather than a peculiarity of the matrix model studied here.

The paper is organized as follows.  Section~\ref{sec:formalism} isolates the
low-moment solution space and reviews multicut factorization and filling
fractions.  Section~\ref{sec:rational-probes} compares Hankel and Cauchy Gram
matrices and introduces the spectral-resolution $\eta$ flow.  Section
\ref{sec:moment_space} defines the filling-fraction manifold and derives the
dressed-saddle conormal response.  Sections
\ref{sec:quartic-diagonal-pick}--\ref{sec:sextic} develop the quartic and
sextic examples.  Section~\ref{sec:conclusion} summarizes the results.

\section{Planar loop equations and spectral geometry}
\label{sec:formalism}

We begin by separating the algebraic freedom left by the planar loop
equation from the additional conditions that select a physical multicut
solution.  The distinction is useful throughout: the loop equation defines
an ambient family of algebraic resolvents, fixed-topology factorization
selects a lower-dimensional locus, and positivity and period conditions
then determine which points on that locus are physical.  This hierarchy
identifies the space in which the geometric construction of
Section~\ref{sec:moment_space} will take place.

\subsection{Loop equation and low-moment space}
\label{subsec:planar-loop}

At fixed potential couplings, \eqref{eq:loop-general} recursively determines
the higher moments once the low moments
\begin{equation}
    \mathbf m=(m_1,\ldots,m_{d-2})
\end{equation}
have been specified.  Before a cut topology is selected, these moments are
therefore local coordinates on a $(d-2)$-dimensional space of algebraic
loop-equation solutions.  This coordinate statement is local: the algebraic
solution space may be branched or singular globally.  Moreover, solving the
loop equation is only the first step.  A physical resolvent must obey the
large-$z$ normalization
\begin{equation}
    W(z)\sim \frac1z,
\end{equation}
be analytic away from its real spectral support, and be the Stieltjes
transform of a positive real measure.  We next impose a regular cut topology
and identify the continuous moduli that survive this local algebraic
restriction.

\subsection{Cut topology, dressed saddles, and filling fractions}
\label{subsec:standard-cut-counting}

Since the discriminant $D$ is a polynomial of degree $2(d-1)$, a regular
$s$-cut solution is characterized by the factorization
\begin{equation}
    D(z)
    =
    M_{d-s-1}(z)^2\,\sigma_{2s}(z),
    \qquad
    \sigma_{2s}(z)=\prod_{\alpha=1}^{2s}(z-e_\alpha).
    \label{eq:standard-s-cut-factorization}
\end{equation}
The two factors have distinct geometric roles.  The $2s$ simple real zeros
$e_\alpha$ are branch points and determine the cut endpoints.  By contrast,
the zeros $r_a$ of $M_{d-s-1}$ are even-multiplicity zeros of $D$ and hence
nonbranching zeros of $y$; regularity requires that they do not coincide
with the branch points.  Their interpretation is especially direct on a
real component of the resolvent set, where
\begin{equation}
    V_{\mathrm{eff}}'(x)
    =
    V'(x)-2W(x)
    =
    y(x).
    \label{eq:intro-effective-potential}
\end{equation}
Consequently, a real zero $r_a$ is a stationary point of the effective
potential felt by a test eigenvalue.  This is the origin of the term \textit{dressed
saddle.}

Having separated the branch points from the nonbranching zeros, we can count
the remaining continuous moduli.  The $2s$ endpoints together with the
$d-s$ coefficients of
$M_{d-s-1}$ give $d+s$ spectral parameters.  The large-$z$ condition
\begin{equation}
    y(z)
    =
    V'(z)-\frac2z+O(z^{-2}),
    \qquad
    z\to\infty,
    \label{eq:standard-s-cut-asymptotics}
\end{equation}
provides $d+1$ independent constraints.  The regular $s$-cut ansatz
therefore leaves
\begin{equation}
    (d+s)-(d+1)=s-1
    \label{eq:standard-filling-count}
\end{equation}
continuous parameters.  They are the independent filling fractions.  Thus a
one-cut curve is fixed locally by its topology and large-$z$ normalization,
whereas a multicut topology defines a family until its fillings are
specified.  For the equilibrium matrix integral, the remaining moduli are
selected by the usual period conditions
\cite{DiFrancesco1995,Eynard2015}.

The same count can be made directly in low-moment coordinates.  The
polynomial $M_{d-s-1}$ has $d-s-1$ zeros, counted with multiplicity.  Before
additional discrete symmetries are imposed, preserving these nonbranching
zeros supplies $d-s-1$ real transverse conditions on the $d-2$ low moments:
a real zero contributes one real condition, while a complex conjugate pair
contributes two.  The fixed-topology family consequently has $s-1$
directions, in agreement with the filling-fraction count above.
Section~\ref{sec:moment_space} turns this counting statement into an explicit
tangent--conormal geometry.

\section{Positivity and exact Cauchy probes}
\label{sec:rational-probes}

The loop equation and cut factorization describe algebraic candidate
resolvents, but they do not ensure that a candidate is the Stieltjes transform
of a positive measure.  We now compare two Gram-matrix organizations of this
positivity information.  The Hankel hierarchy uses polynomial probes at
infinity; the Pick matrix uses Cauchy kernels at finite spectral points.  A
matched-information comparison will separate a change of basis from the
genuinely finite-plane information of the exact rational transform.

\subsection{Hankel positivity}

Let
\begin{equation}
    p(x)=\sum_{k=0}^{K}c_kx^k,
\end{equation}
be an arbitrary polynomial of degree at most $K$.  Positivity of the
eigenvalue measure requires
\begin{equation}
    \int_{\mathbb R}|p(x)|^2 \rho(x)\,\dd x
    =
    \sum_{i,j=0}^{K}\bar c_i c_j\,m_{i+j}
    \geq0.
\end{equation}
Since the coefficients $c_k$ are arbitrary, the corresponding Hankel matrix
\begin{equation}
    H^{(K)}_{ij}=m_{i+j} = \int_{\mathbb R} x^{i+j}\rho(x)\,\dd x,
    \qquad
    i,j=0,\ldots,K,
    \label{eq:hankel}
\end{equation}
must be positive semidefinite:
\begin{equation}
    H^{(K)}\succeq0.
    \label{eq:hankel-psd}
\end{equation}
Combining this hierarchy with the loop equations gives the standard moment
bootstrap \cite{Lin2020, AndersonKruczenski2017,KazakovZheng2022}.  At any finite $K$,
\eqref{eq:hankel-psd} is only a truncated necessary condition.  The full
moment problem requires positivity at every order, supplemented by the
appropriate growth or support conditions.
 
\subsection{Pick positivity and the Cauchy Gram matrix}
\label{sec:diagonal-pick-herglotz}

Let
\begin{equation}
    f_{z_i}(x)=\frac{1}{z_i-x},
    \qquad
    z_i\in\HH_+,
\end{equation}
be a collection of Cauchy kernels.  With respect to the spectral
measure $\rho(x)\,\dd x$, their Gram matrix is
\begin{equation}
    P_{ij}
    \equiv
    \langle f_{z_i},f_{z_j}\rangle_\rho
    =
    \int_{\mathbb R}
    \frac{\rho(x)\,\dd x}
         {(\overline{z_i}-x)(z_j-x)}.
    \label{eq:pick-gram}
\end{equation}
For arbitrary coefficients $c_i$,
\begin{equation}
    \sum_{i,j}\overline{c_i}P_{ij}c_j
    =
    \int_{\mathbb R}\rho(x)
    \left|
        \sum_i\frac{c_i}{z_i-x}
    \right|^2\dd x
    \geq0,
\end{equation}
and hence
\begin{equation}
    P\succeq0.
    \label{eq:pick-psd}
\end{equation}
Positivity of the Pick matrix is therefore an immediate consequence of
positivity of the underlying spectral measure.

The same matrix can be expressed directly in terms of the resolvent.  Using
\begin{equation}
    W(z)
    =
    \int_{\mathbb R}\frac{\rho(x)}{z-x}\,\dd x,
\end{equation}
one obtains
\begin{equation}
    P_{ij}
    =
    \frac{\overline{W(z_i)}-W(z_j)}
         {z_j-\overline{z_i}}.
    \label{eq:pick-definition}
\end{equation}
Thus the Pick matrix packages finite-plane values of the resolvent into the
Gram matrix of the corresponding Cauchy probes.

For a single node, \eqref{eq:pick-gram} reduces to
\begin{equation}
    P(z,z)
    =
    -\frac{\operatorname{Im}W(z)}
           {\operatorname{Im}z}
    =
    \int_{\mathbb R}
    \frac{\rho(x)\,\dd x}{|z-x|^2}
    \geq0.
    \label{eq:pick-diagonal-general}
\end{equation}
In particular, for $z=E+\ii\eta$ with $\eta>0$,
\begin{equation}
    \operatorname{Im}W(E+\ii\eta)
    =
    -\eta
    \int_{\mathbb R}
    \frac{\rho(x)}
         {(E-x)^2+\eta^2}\,\dd x
    \leq0.
    \label{eq:herglotz-sign}
\end{equation}
Consequently, $W$ is a Stieltjes transform and
\begin{equation}
    F=-W:\HH_+\mapsto\HH_+
\end{equation}
is a
Herglotz--Nevanlinna, or Pick, function
\cite{Pick1917,Donoghue1974,GesztesyTsekanovskii2000,Nedic2021}.

We refer to $P(z,z)$ as the \emph{Pick response}.   The
off-diagonal entries become relevant when two or more nodes are considered:
they test whether the responses at different spectral positions and
resolutions are compatible with the same positive measure.  For example,
positivity of a two-node principal minor requires
\begin{equation}
    |P_{12}|^2\leq P_{11}P_{22}.
\end{equation}

A Pick matrix constructed from one fixed finite node set therefore supplies
only a necessary positivity test for a candidate algebraic resolvent.
Requiring $P\succeq0$ for every finite collection of upper-half-plane nodes
gives the full Pick condition and, together with the Stieltjes
normalization at infinity, characterizes a common positive spectral measure.

The Herglotz representation, Pick interpolation problem, and rational moment
problem are classical
\cite{Akhiezer1965,KreinNudelman1977,Pick1917,Donoghue1974,
AglerMcCarthy2002,GesztesyTsekanovskii2000,Nedic2021,Almendral2003,
NailwalZalar2025}.  Related Herglotz and Nevanlinna constructions have also
been used in spectral reconstruction problems
\cite{FeiYehGull2021,FeiYehZgidGull2021,HuangGullLin2023,
BergamaschiJayOare2023,FieldsChrist2026,AbbottJayOare2026,
AbbottFieldsJayOareSaccardi2026}.

\subsection{Information-matched comparison with the Hankel bootstrap}
\label{subsec:information-matched}

Having defined both Gram matrices, we now compare them at fixed finite
information.  This step is necessary before assigning any additional
strength to the rational formulation.  Begin with the polynomial basis
\begin{equation}
    e_n(x)=x^n,
    \qquad
    n=0,\ldots,K,
\end{equation}
and truncate the large-$z$ expansion of each Cauchy kernel to that same
space:
\begin{equation}
    f^{(K)}_{z_i}(x)
    =
    \sum_{n=0}^{K}\frac{x^n}{z_i^{\,n+1}}.
    \label{eq:truncated-cauchy}
\end{equation}
Writing the associated change-of-basis matrix as
\begin{equation}
    C_{in}=\frac{1}{z_i^{\,n+1}},
\end{equation}
the Gram matrix of the truncated probes becomes
\begin{equation}
    P^{(K)}=\overline C\,H^{(K)}\,C^{\mathsf T}.
    \label{eq:congruence}
\end{equation}
For $K+1$ distinct nodes in the upper half-plane, $C$ is a scaled
Vandermonde matrix in $1/z_i$ and is therefore invertible.  Consequently,
\begin{equation}
    P^{(K)}\succeq0
    \quad\Longleftrightarrow\quad
    H^{(K)}\succeq0.
    \label{eq:equivalent-psd}
\end{equation}
Thus, at matched finite polynomial information, the rational and monomial
bases define exactly the same positivity cone.

The exact rational construction begins only when the truncation is removed.
The full kernel is
\begin{equation}
    \frac{1}{z-x}
    =
    \frac1z+\frac{x}{z^2}+\frac{x^2}{z^3}+\cdots.
    \label{eq:resummation}
\end{equation}
As a series in $x/z$, this expansion converges only when the pole lies
outside the spectral radius.  The exact Cauchy transform, however, is
defined for every $z$ away from the real support.  In the algebraic
matrix-model problem, it can be evaluated at finite $z$ by solving the exact
loop equation and continuing the physical branch from infinity.  The
finite-plane localization therefore belongs to the exact rational probe and
the resummed resolvent; it is not a consequence of a finite change of basis.

This distinction suggests a direct finite-node bootstrap of the
undetermined low moments.  One inserts the algebraic loop-equation solution
\begin{equation}
    W(z;\mathbf m)
    =
    \frac12\left[V'(z)-\sqrt{D(z;\mathbf m)}\right]
\end{equation}
into the Pick matrix
\begin{equation}
    P_{ij}(\mathbf m)
    =
    \frac{\overline{W(z_i;\mathbf m)}-W(z_j;\mathbf m)}
         {z_j-\overline{z_i}},
\end{equation}
and excludes trial moments for which
$P(\mathbf m)\not\succeq0$.  In this sense, the construction is a
rational finite-plane bootstrap analogous to the Hankel hierarchy.

For numerical bounds, the two formulations require a careful comparison.
Hankel positivity is an affine semidefinite constraint when the moments
entering the matrix are retained as independent variables. The planar loop
equations, however, contain products of moments; after higher moments are
eliminated, the constraints on the independent low moments are generally
nonlinear. For example, the quartic recursion gives
$m_6=(m_1^2+2m_2)/g+(m_2+1)/g^2$. Convex semidefinite formulations
therefore require an appropriate treatment of the nonlinear loop equations,
such as the relaxation of \cite{KazakovZheng2022}.

The exact Pick formulation introduces the additional need to evaluate the
algebraic resolvent on a consistently continued sheet. A pointwise
principal-square-root choice can jump between sheets as discriminant zeros
move, producing discontinuous constraints and spurious positivity failures.
The strength and conditioning of a finite test also depend on the relative
positions of its nodes and the spectral support. Appendix~\ref{appendixA}
examines these issues for the asymmetric quartic, comparing specified
Hankel cutoffs with exact Pick tests and checking continuation, node
placement, and eigenvalue tolerances.

These are properties of the finite-node algebraic implementation. The full
Pick--Nevanlinna characterization, with Stieltjes normalization, expresses
positivity of a common spectral measure. We use exact Cauchy kernels
primarily to measure local spectral geometry; their numerical bootstrap
provides a complementary organization of positivity information.

\subsection{Spectral resolution and the \texorpdfstring{$\eta$}{eta} flow}
\label{subsec:spectral-resolution}

We now make the localization scale explicit.  For a probe at
$z=E+\ii\eta$, define the finite-resolution density
\begin{equation}
    \rho_\eta(E)
    \equiv
    -\frac1\pi\operatorname{Im}W(E+\ii\eta)
    =
    \frac{\eta}{\pi}
    P(E+\ii\eta,E+\ii\eta).
    \label{eq:finite-resolution-density}
\end{equation}
Equation~\eqref{eq:herglotz-sign} then gives
\begin{equation}
    \rho_\eta(E)
    =
    \frac1\pi
    \int_{\mathbb R}\rho(x)
    \frac{\eta\,\dd x}{(E-x)^2+\eta^2}
    =
    (K_\eta*\rho)(E),
    \qquad
    K_\eta(E)
    =
    \frac1\pi\frac{\eta}{E^2+\eta^2}.
    \label{eq:poisson-regularization}
\end{equation}
Thus $\rho_\eta$ is the Poisson extension of the physical density.  The
parameter $E$ fixes the center of the probe, while $\eta$ provides a
continuous resolution scale.  For a finite-$N$ microscopic density, this
convolution replaces each eigenvalue by a Lorentzian of full width at half
maximum $2\eta$.

Changing the resolution is governed by the Poisson semigroup:
\begin{equation}
    K_{\eta_1}*K_{\eta_2}=K_{\eta_1+\eta_2},
    \qquad
    \rho_{\eta_1+\eta_2}
    =
    K_{\eta_1}*\rho_{\eta_2}.
    \label{eq:eta-semigroup}
\end{equation}
Equivalently, Fourier transformation gives
\begin{equation}
    \widehat{\rho_\eta}(t)
    =
    e^{-\eta|t|}\widehat{\rho}(t),
\end{equation}
and hence
\begin{equation}
    \partial_\eta\rho_\eta(E)
    =
    -|\partial_E|\rho_\eta(E).
\end{equation}
Here $|\partial_E|$ is the Fourier multiplier with symbol $|t|$.  This is
the classical Poisson semigroup.  Its generator
$-|\partial_E|=-(-\partial_E^2)^{1/2}$ is the negative of the
Dirichlet-to-Neumann operator for harmonic extension to the upper
half-plane, with the outward-normal convention~\cite{CaffarelliSilvestre2007}.  We refer to this evolution as the
\emph{$\eta$ flow}.  It is a flow of spectral resolution, rather than an RG
flow: increasing $\eta$ suppresses short-distance spectral information,
whereas decreasing $\eta$ resolves progressively finer structure.  The
analyticity of $W$ in the upper half-plane also implies that
$\rho_\eta(E)$ is harmonic:
\begin{equation}
    (\partial_E^2+\partial_\eta^2)\rho_\eta(E)=0.
\end{equation}

The boundary behavior of this harmonic extension distinguishes the local
spectral regimes.  At a continuity point in the bulk,
\begin{equation}
    \rho_\eta(E)\longrightarrow\rho(E).
\end{equation}
If $E_\star$ is an edge with
$\rho(E_\star-s)\sim Cs^\alpha$, $-1<\alpha<1$, then
\begin{equation}
    \rho_\eta(E_\star)
    \sim
    \frac{C}{2\cos(\pi\alpha/2)}\,\eta^\alpha.
    \label{eq:edge-scaling}
\end{equation}
In particular, a regular square-root edge gives
$\rho_\eta(E_\star)=O(\eta^{1/2})$.  By contrast, if $E$ lies strictly
inside a gap, analyticity across the real axis gives
\begin{equation}
    \rho_\eta(E)
    =
    \frac{\eta}{\pi}
    \int_{\mathbb R}\frac{\rho(x)}{(E-x)^2}\,\dd x
    +O(\eta^3),
    \label{eq:gap-small-eta}
\end{equation}
so the signal vanishes linearly.  The small-$\eta$ scaling of the same
probe family therefore distinguishes bulk points, edges, and gaps without
requiring a global reconstruction of the density.

The opposite, large-$\eta$ regime connects the exact rational description
back to the moment expansion.  For an even density,
\begin{equation}
    W(\ii\eta)
    =
    -\ii\eta
    \int_{\mathbb R}\frac{\rho(x)}{x^2+\eta^2}\,\dd x,
\end{equation}
and therefore obeys the exact identity
\begin{equation}
    -W(\ii\eta)
    =
    \ii\eta\,P(\ii\eta,\ii\eta).
    \label{eq:W-P-imaginary-axis}
\end{equation}
Its expansion at infinity is
\begin{equation}
    -W(\ii\eta)
    =
    \frac{\ii}{\eta}
    -\frac{\ii m_2}{\eta^3}
    +\frac{\ii m_4}{\eta^5}
    -\cdots,
\end{equation}
which yields
\begin{equation}
    P(\ii\eta,\ii\eta)
    =
    \frac1{\eta^2}
    -\frac{m_2}{\eta^4}
    +\frac{m_4}{\eta^6}
    -\cdots.
    \label{eq:momentexpansion}
\end{equation}
Thus the moment expansion of $-W(\ii\eta)$ reproduces the large-$\eta$
expansion of the Pick response.  The distinction between the two
descriptions is important: the exact identity
\eqref{eq:W-P-imaginary-axis} holds for every $\eta>0$ on an even physical
branch, while the moment series converges only outside the spectral radius.
The exact Cauchy probe can therefore enter the local finite-plane regime in
which an expansion about infinity is no longer the natural description.

\section{Moment-space geometry and dressed-saddle response}
\label{sec:moment_space}

We now combine the algebraic counting of Section~\ref{sec:formalism} with the
finite-resolution probes of Section~\ref{sec:rational-probes}.  We first
define the fixed-topology locus and derive its tangent--conormal geometry
directly from the double-zero constraints.  Only after this geometric object
has been identified do we show that the small-resolution Pick response
measures its conormal directions.

\subsection{The filling-fraction manifold and its conormals}
\label{subsec:filling-manifold}

Fix the potential couplings.  The planar loop equation then leaves the
$d-2$ low moments
\begin{equation}
    \mathbf m=(m_1,\ldots,m_{d-2})
\end{equation}
undetermined. Locally, one can  define the ambient low-moment space
\begin{equation}
    \mathcal M\simeq\mathbb R^{d-2}.
\end{equation}
Within this ambient space, let
\begin{equation}
    \mathcal F_s\subset\mathcal M
\end{equation}
denote the local family of regular planar solutions with a fixed $s$-cut
topology, before the equilibrium period conditions are imposed.  Along $\mathcal F_s$, the discriminant $D(z,\mathbf m)$ has $2s$ simple real
zeros corresponding to the branch points and $d-s-1$ nonbranching double
zeros. The remaining moduli are
parametrized by the independent filling fractions, and we therefore refer to
$\mathcal F_s$ as the \emph{filling-fraction manifold}.

\paragraph{Local dimension.}

To determine the dimension of this manifold, recall the
regular factorization
\begin{equation}
    D(z;\mathbf m)
    =
    M_{d-s-1}(z)^2
    \prod_{\alpha=1}^{2s}(z-e_\alpha).
\end{equation}
Since $M_{d-s-1}$ has degree $d-s-1$, its zeros provide $d-s-1$
nonbranching double-zero conditions on the low moments.   For a real double zero $z=r$,
\begin{equation}
    D(r;\mathbf m)=0,
    \qquad
    D_z(r;\mathbf m)=0
\end{equation}
are two real equations, but $r$ is itself one additional real unknown. Therefore we have  one real transverse condition.  Similarly, a
complex  conjugate pair contributes two real transverse conditions once the
real and imaginary parts of its position are allowed to vary.  Assuming
that these constraints are independent, we have
\begin{equation}
    \operatorname{codim}_{\mathcal M}\mathcal F_s=d-s-1,
\end{equation}
and hence
\begin{equation}
    \dim\mathcal F_s
    =
    (d-2)-(d-s-1)
    =
    s-1.
    \label{eq:dressed-saddle-dimension-count}
\end{equation}
This reproduces the standard multicut count in low-moment coordinates: a
regular $s$-cut family has precisely $s-1$ local moduli, namely the
independent filling fractions.  The dimension count is not itself a new
solution method.  Its role is to identify the tangent and transverse spaces
that the finite-plane Pick response will measure.

\paragraph{Conormal covectors.}

We next differentiate the double-zero constraints to extract their
transverse covectors.
Let $r_a$ be a regular nonbranching double zero of a curve in
$\mathcal F_s$, so that
\begin{equation}
    D(r_a;\mathbf m)=0,
    \qquad
    D_z(r_a;\mathbf m)=0,
    \qquad
    D_{zz}(r_a;\mathbf m)\neq0.
    \label{eq:dressed-saddle-regularity}
\end{equation}
Now choose a differentiable curve of fixed-topology solutions,
\begin{equation}
    t\longmapsto\mathbf m(t)\in\mathcal F_s
\end{equation}
and let $r_a(t)$ denote the corresponding motion of the double zero.
Differentiating the first condition in
\eqref{eq:dressed-saddle-regularity} yields
\begin{equation}
    \nabla_{\mathbf m}D(r_a;\mathbf m)
       \cdot\delta\mathbf m
    +
    D_z(r_a;\mathbf m)\,\delta r_a
    =
    0,
\end{equation}
where
\begin{equation}
    \delta\mathbf m
    =
    \left.\frac{d\mathbf m(t)}{dt}\right|_{t=0}\delta t
    \in T_{\mathbf m}\mathcal F_s.
\end{equation}
At a double zero $D_z(r_a;\mathbf m)=0$.  The displacement $\delta r_a$
therefore drops out, and every tangent variation obeys
\begin{equation}
    \nabla_{\mathbf m}D(r_a;\mathbf m)
       \cdot\delta\mathbf m
    =
    0.
\end{equation}
The derivative of the constraint consequently defines
\begin{equation}
    \boldsymbol n_a
    \equiv
    \boldsymbol n(r_a)
    =
    -\frac14\,\nabla_{\mathbf m}D(r_a;\mathbf m)
    =
    \nabla_{\mathbf m}\mathcal Q(r_a;\mathbf m),
    \qquad
    \boldsymbol n_a\cdot\delta\mathbf m=0.
    \label{eq:general-conormal-tangent}
\end{equation}
The last equality uses the fixed-coupling identity
\begin{equation}
    D(z;\mathbf m)=V'(z)^2-4\mathcal Q(z;\mathbf m).
\end{equation}
Here $\nabla_{\mathbf m}D$ is the coordinate row of derivatives with
respect to the low moments.  Its pairing with $\delta\mathbf m$ is the
natural covector-vector pairing.  Thus $\boldsymbol n_a$ is intrinsically a
\emph{conormal covector}: it is the derivative of a constraint and
annihilates $T_{\mathbf m}\mathcal F_s$.  Only after a metric is chosen on
the ambient moment space can one use that metric to identify this covector
with an ordinary normal vector.  The dot in
\eqref{eq:general-conormal-tangent} denotes the natural pairing and does
not assume such a metric.

The preceding argument identifies the conormal supplied by a single saddle.
A real dressed saddle supplies one real covector.  A complex saddle
$z_a$ occurs together with its complex conjugate and supplies two real
covectors,
\begin{equation}
    \operatorname{Re}\boldsymbol n(z_a),
    \qquad
    \operatorname{Im}\boldsymbol n(z_a).
\end{equation}
At a generic point of the regular locus, the resulting $d-s-1$ real
covectors are independent.  Since this number equals the codimension found
above, they span the full conormal space:
\begin{equation}
    N_{\mathbf m}^{*}\mathcal F_s
    =
    \operatorname{span}_{\mathbb R}
    \{\boldsymbol n_1,\ldots,\boldsymbol n_{d-s-1}\},
    \qquad
    \dim N_{\mathbf m}^{*}\mathcal F_s=d-s-1.
    \label{eq:filling-conormal-space}
\end{equation}
Here and below, the real and imaginary parts of a nonreal conjugate pair are
counted separately.  Each $\boldsymbol n_a$ has $d-2$ components because it lies
in the ambient cotangent space $T_{\mathbf m}^{*}\mathcal M$; the number of
independent conormal directions is nevertheless only $d-s-1$.

\paragraph{The common kernel.}

Because each conormal annihilates every tangent vector, the tangent space is
contained in their common kernel.  Independence of the $d-s-1$ conormals
makes this kernel $(s-1)$-dimensional, equal to
$\dim\mathcal F_s$ in \eqref{eq:dressed-saddle-dimension-count}.  The
containment is therefore an equality:
\begin{equation}
    T_{\mathbf m}\mathcal F_s
    =
    \bigcap_{a=1}^{d-s-1}\ker \boldsymbol n_a,
    \qquad
    \dim T_{\mathbf m}\mathcal F_s=s-1.
    \label{eq:filling-tangent-kernel}
\end{equation}
Equivalently, assemble the conormals as the rows of the matrix
\begin{equation}
    N=
    \begin{pmatrix}
        \boldsymbol n_1\\
        \vdots\\
        \boldsymbol n_{d-s-1}
    \end{pmatrix}
    \in
    \mathbb R^{(d-s-1)\times(d-2)}.
\end{equation}
The tangent--conormal duality then takes the compact form
\begin{equation}
    N_{\mathbf m}^{*}\mathcal F_s=\operatorname{row}(N),
    \qquad
    T_{\mathbf m}\mathcal F_s=\ker N.
\end{equation}
Thus the local double-zero conditions remove the transverse moment
directions while leaving the $s-1$ filling-fraction directions
undetermined. 

\paragraph{Critical locus}

This argument applies only on the regular locus.  Suppose instead that a
zero of $M_{d-s-1}$ has multiplicity $k>1$.  It is then a nongeneric dressed
saddle for which
\begin{equation}
    D(z)\sim(z-r)^{2k}.
\end{equation}
Preserving this degeneracy requires additional conditions.  Such points lie
on higher-codimension critical loci, where the independence and dimension
assumptions used above need not hold.

\subsection{Pick response near a real dressed saddle}
\label{subsec:cauchy-conormal}

We now relate these already-defined conormal covectors to a directly
evaluable finite-plane response.  At fixed potential, define the low-moment
derivative of the loop polynomial by
\begin{equation}
    \boldsymbol n(z)
    \equiv
    -\frac14\nabla_{\mathbf m}D(z)
    =
    \nabla_{\mathbf m}\mathcal Q(z).
\end{equation}
Differentiating the algebraic resolvent with respect to the low moments gives
\begin{equation}
    \nabla_{\mathbf m}W(z)
    =
    -\frac{1}{4y(z)}
    \nabla_{\mathbf m}D(z)
    =
    \frac{\boldsymbol n(z)}{y(z)}.
    \label{eq:general-dW-dm}
\end{equation}
\paragraph{A single real saddle.}

Let $r$ be a regular real dressed saddle lying in a real component of the
resolvent set.  Locally,
\begin{equation}
    y(z)
    =
    a_r(z-r)+O((z-r)^2),
    \qquad
    a_r\neq0.
    \label{eq:general-real-dressed-saddle}
\end{equation}
A probe $z=E+\ii\eta$ with $E-r=O(\eta)$ then has the response
\begin{equation}
    \nabla_{\mathbf m}
    P(E+\ii\eta,E+\ii\eta)
    =
    \frac{1}{a_r}
    \frac{\boldsymbol n(r)}{(E-r)^2+\eta^2}
    +O(1).
    \label{eq:general-cauchy-lorentzian}
\end{equation}
Centering the probe at the dressed saddle gives the special case
\begin{equation}
    \nabla_{\mathbf m}
    P(r+\ii\eta,r+\ii\eta)
    =
    \frac{1}{a_r\eta^2}\,\boldsymbol n(r)
    +O(1),
    \qquad
    \eta\to0^+.
    \label{eq:general-cauchy-conormal}
\end{equation}
The response is therefore parametrically enhanced precisely along the
conormal to the filling-fraction manifold.  If $\delta\mathbf m$ is tangent
to $\mathcal F_s$, then $\boldsymbol n(r)\cdot\delta\mathbf m=0$, and the leading
$\eta^{-2}$ term cancels in the directional response.

\paragraph{Several real saddles.}

To reconstruct more than one conormal direction, consider independent real
dressed saddles $r_i$ and define the response matrix
\begin{equation}
    \mathcal R_{ia}(\eta)
    =
    \frac{\partial}{\partial m_a}
    P(r_i+\ii\eta,r_i+\ii\eta).
\end{equation}
Let $N$ denote the matrix whose rows are $\boldsymbol n(r_i)^T$.  Equation
\eqref{eq:general-cauchy-conormal} then implies
\begin{equation}
    \eta^2\mathcal R(\eta)
    \longrightarrow
    \operatorname{diag}
    (a_{r_1}^{-1},\ldots,a_{r_k}^{-1})\,N .
    \label{eq:general-response-matrix-limit}
\end{equation}
If these conormals are independent, the limiting response has the same
kernel as $N$:
\begin{equation}
    \ker\!\left[
       \lim_{\eta\to0^+}\eta^2\mathcal R(\eta)
    \right]
    =
    \ker N
    =
    T_{\mathbf m}\mathcal F_s .
    \label{eq:general-response-kernel}
\end{equation}
This is the central differential statement of the rational-probe
construction.  The row space of the rescaled small-resolution response
reconstructs the local conormal space of the fixed-topology family, while
its kernel reconstructs the tangent space.

\paragraph{Nonreal saddles.}

The same mechanism applies to a dressed saddle at a nonreal point
$z_a\in\HH_+$, but the appropriate limit is now taken directly in the
finite complex plane.  Since
$y(z)\sim a_a(z-z_a)$,
\begin{equation}
    \nabla_{\mathbf m}W(z)
    \sim
    \frac{\boldsymbol n(z_a)}{a_a(z-z_a)}.
\end{equation}
The susceptibility becomes singular as $z$ approaches $z_a$.  There is no
one-dimensional $\eta\to0$ boundary limit associated with such a saddle;
the natural probe instead approaches the finite complex point itself.

\subsection{Division of roles: probe, positivity, reality, and analyticity}
\label{sec:probe-positivity}

Before turning to the examples, it is useful to separate the inputs used by
the construction.  The finite-plane response is not a new positivity
principle: localization is already a property of the Cauchy kernel.  Even
for a signed trial density,
\begin{equation}
    -\frac1\pi\operatorname{Im}W(E+\ii\eta)
    =
    \int_{\mathbb R}\rho(x)K_\eta(E-x)\,\dd x
\end{equation}
remains a localized finite-resolution transform.  Positivity enters only
when one asks whether the trial resolvent can arise from a single
nonnegative measure.

At one node, this requirement is exactly the Pick sign
\begin{equation}
    P(z,z)\geq0.
\end{equation}
With several nodes, $P\succeq0$ correlates different positions and
resolutions and tests whether their local responses are compatible with the
same positive spectrum.  This cross-node requirement is the content of Pick
positivity.  Rationality plays a different role: it provides the movable
finite-plane pole and the continuously tunable resolution.

The examples also use two constraints that must remain logically separate
from positivity.  First, symmetry or a real spectral gap may force $W$ or
$y$ to have a definite phase on a special locus.  The corresponding
condition on $D=y^2$ produces \emph{reality inequalities} and their
envelopes without invoking the Pick sign.  Second, the physical resolvent
must be analytic throughout the upper half-plane.  A simple zero of $D$ in
that domain would create an unphysical branch point, while any
nonbranching zero must have even multiplicity.  Finally, the algebraic
square-root sheet is fixed globally by
\begin{equation}
    W(z)\sim\frac1z,
    \qquad
    z\to\infty,
\end{equation}
and must be continued consistently through every nonbranching zero.

The resulting division of roles is
\begin{equation}
    \begin{aligned}
    \text{Cauchy kernel}
    &\;&\longrightarrow\;&
    \text{localization and resolution},\\
    \text{diagonal Pick sign}
    &\;&\longrightarrow\;&
    \text{one-node positivity},\\
    \text{full Pick matrix}
    &\;&\longrightarrow\;&
    \text{cross-node consistency},\\
    \text{reality on special loci}
    &\;&\longrightarrow\;&
    \text{moment-space envelopes},\\
    \text{analyticity}
    &\;&\longrightarrow\;&
    \text{exclusion of nonreal branch points},\\
    \text{branch continuation}
    &\;&\longrightarrow\;&
    \text{the physical algebraic sheet}.
    \end{aligned}
    \label{eq:division-of-roles}
\end{equation}
For a multicut solution, these local constraints determine the transverse
geometry of the filling-fraction manifold.  The global period conditions
are still required to fix the remaining longitudinal directions.

\section{Symmetric quartic calibrations}
\label{sec:quartic-diagonal-pick}

The symmetric quartic provides a controlled setting in which the physical
moments and the complete factorization of the spectral curve are known.  Its
purpose here is therefore diagnostic: it separates the roles of reality,
Pick positivity, analyticity, and continuation of the algebraic sheet
before these ingredients are applied to families with genuine filling
moduli.  The detailed envelope optimizations, branch analysis, and
factorizations are collected in
Appendix~\ref{app:symmetric-quartic-details}.

Consider
\begin{equation}
    V(x)=\frac{\mu}{2}x^2+\frac{g}{4}x^4,
    \qquad \mu=\pm1,
    \label{eq:quartic-main-potential}
\end{equation}
on the $\mathbb Z_2$-symmetric planar branch, for which
$m_{2k+1}=0$.  The loop equation depends on the single undetermined moment
$m_2$, and the discriminant is
\begin{equation}
    \Delta_{\mu,g}(z)
    =g^2z^6+2\mu g z^4+(1-4g)z^2-4\mu-4gm_2.
    \label{eq:quartic-main-discriminant}
\end{equation}
Reality of the symmetric resolvent implies
$W(\ii\eta)\in\ii\mathbb R$ and hence
$\Delta_{\mu,g}(\ii\eta)\leq0$.  With $q=\eta^2$, this becomes the
one-parameter envelope
\begin{equation}
    m_2\geq L_{\mu,g}(q),
    \qquad
    L_{\mu,g}(q)
    =-\frac{\mu}{g}-\frac{g}{4}q^3
      +\frac{\mu}{2}q^2-\frac{1-4g}{4g}q,
    \label{eq:quartic-main-envelope}
\end{equation}
when $g>0$.  This is a reality condition, not yet the Pick sign.  The
orientation of the physical branch supplies the complementary information
and, in the regimes described below, forces saturation of the optimized
envelope at a dressed saddle.

\subsection{The stable branch and migration of the dressed saddle}

For $\mu=+1$ and $g>0$, maximizing
\eqref{eq:quartic-main-envelope} gives
\begin{equation}
    q_*=\frac{2+\sqrt{1+12g}}{3g},
    \qquad
    m_2=\mathcal B_+(g)
    =\frac{(1+12g)^{3/2}-18g-1}{54g^2}.
    \label{eq:quartic-main-stable-result}
\end{equation}
The reason the reality bound is saturated is global.  Writing
$P_\eta=P(\ii\eta,\ii\eta)>0$, one has
\begin{equation}
    y(\ii\eta)
    =\ii\eta\bigl[1-g\eta^2+2P_\eta\bigr].
\end{equation}
The bracket is positive near the real axis and negative at large $\eta$.
Continuation of the physical sheet therefore forces
$y(\ii\kappa)=0$ for some $\kappa>0$.  At the optimized point,
\begin{equation}
    \kappa^2=q_*,
    \qquad
    \Delta_{+,g}(\ii\kappa)
    =\Delta_{+,g}'(\ii\kappa)=0.
\end{equation}
The reality envelope is thus tangent to the physical solution precisely
where the imaginary-axis probe reaches a regular dressed saddle.  In the
Gaussian limit, $q_*\sim g^{-1}\to\infty$ and
$m_2=1-2g+O(g^2)$.

The same algebraic expression for $m_2$ continues formally to
$-1/12<g<0$, but the geometry changes. The real-axis quartic integral
is not convergent in this range; we consider its formal one-cut
continuation, whose compact density remains nonnegative in the stated interval.  In that range
\begin{equation}
    y(\ii\eta)
    =\ii\eta\bigl[1+|g|\eta^2+2P_\eta\bigr]
\end{equation}
has no zero on the positive imaginary axis.  The required sign change occurs
instead on the real exterior of the cut, where a regular dressed saddle lies
at
\begin{equation}
    r^2
    =\frac{-2-\sqrt{1+12g}}{3g}.
    \label{eq:quartic-main-negative-saddle}
\end{equation}
The distinguished spectral point therefore migrates continuously according
to
\begin{equation}
    g>0: z_*=\pm\ii\kappa,
    \qquad
    g\to0: |z_*|\to\infty,
    \qquad
    -\frac1{12}<g<0: z_*=\pm r.
    \label{eq:quartic-main-migration}
\end{equation}
At $g=-1/12$ the dressed saddles collide with the branch points and
\begin{equation}
    D(z)=\frac1{144}(z^2-8)^3,
\end{equation}
so the regular double-zero geometry degenerates into cubic branch points.

\subsection{The symmetric double well}

For $\mu=-1$ and $0<g<1/4$, the physical symmetric solution has two cuts.
Oddness and analyticity in the central gap imply $W(0)=0$, which immediately
fixes
\begin{equation}
    m_2=\frac1g.
    \label{eq:quartic-main-twocut-result}
\end{equation}
The discriminant factorizes as
\begin{equation}
    D(z)=g^2z^2(z^2-A^2)(z^2-B^2),
    \qquad
    A^2=\frac{1-2\sqrt g}{g},
    \qquad
    B^2=\frac{1+2\sqrt g}{g}.
\end{equation}
The origin is therefore a regular dressed saddle in the central gap.  The
imaginary-axis and real-gap reality inequalities meet at
$m_2=1/g$; in this symmetric benchmark, that two-sided pinch geometrizes a
value already fixed by $W(0)=0$.

At $g=1/4$, the inner endpoints collide with the origin and the
discriminant develops a fourth-order zero.  For $g>1/4$, the support is one
cut and the central real-gap test is no longer available.  The optimized
imaginary-axis envelope instead gives
\begin{equation}
    q_*=\frac{\sqrt{1+12g}-2}{3g},
    \qquad
    m_2=\mathcal B_-(g)
    =\frac{(1+12g)^{3/2}+18g+1}{54g^2}.
    \label{eq:quartic-main-onecut-result}
\end{equation}
Here the physical Pick branch changes orientation between the real axis
and infinity and is forced through the pair
$z=\pm\ii\sqrt{q_*}$.  The resulting zeros are nonbranching only on the
saturated locus.  A transverse displacement splits them into simple zeros
in the upper and lower half-planes and is excluded by analyticity.

These examples establish the logic used later.  Reality produces
moment-space envelopes, the Pick sign fixes the orientation of the
physical branch, and continuation determines where that branch must pass
through a dressed saddle.  The rational probe localizes this spectral event
but does not define a positivity cone stronger than the information-matched
Hankel cone of Section~\ref{subsec:information-matched}.  The asymmetric
quartic is the first case in which the same mechanism determines a conormal
to a nontrivial filling-fraction family rather than a single symmetric
solution.

\section{The asymmetric quartic}
\label{sec:asymmetric-quartic}

The symmetric examples reduce the low-moment space to one dimension and
therefore select isolated solutions once a cut topology is imposed.  We now
remove the equal-filling restriction.  The resulting asymmetric two-cut
family is the first example for which the tangent and conormal spaces are
both nontrivial.

Consider the quartic double-well potential
\begin{equation}
    V(x)=-\frac{x^2}{2}+\frac{g}{4}x^4,
    \qquad 0<g<\frac14,
    \label{eq:asym-potential}
\end{equation}
without imposing equal filling fractions between the two wells. The odd
moments need not vanish, and the planar loop equation is
\begin{equation}
    W(z)^2-\bigl(gz^3-z\bigr)W(z)
    -1+g\bigl(z^2+m_1z+m_2\bigr)=0.
    \label{eq:asym-loop}
\end{equation}
The discriminant is
\begin{equation}
    D(z)=g^2z^6-2gz^4+(1-4g)z^2
    -4gm_1z+4(1-gm_2).
    \label{eq:asym-D}
\end{equation}
The physical sheet is fixed by
\begin{equation}
    W(z)=\frac1z+O(z^{-2}),
    \qquad
    y(z)=V'(z)-\frac2z+O(z^{-2}),
    \label{eq:asym-sheet}
\end{equation}
at infinity.

A regular two-cut quartic has four simple real branch points and, since
$\deg M_{d-s-1}=1$, one regular dressed saddle. Let the support be
\begin{equation}
    [e_1,e_2]\cup[e_3,e_4],
    \qquad e_1<e_2<e_3<e_4,
\end{equation}
and denote the central gap by $G=(e_2,e_3)$. Positivity of the density fixes
the orientation of the physical branch on the cuts,
\begin{equation}
    y_+(x)=2\pi\ii\rho(x),
\end{equation}
so analytic continuation through the two inner endpoints gives
\begin{equation}
    y(e_2+\epsilon)>0,
    \qquad
    y(e_3-\epsilon)<0,
    \qquad 0<\epsilon\ll1.
    \label{eq:asym-gap-signs}
\end{equation}
The real function $y(x)$ must therefore vanish at some point
$r\in(e_2,e_3)$.  In the regular phase this zero of $y$ is simple, and hence
\begin{equation}
    D(r)=D'(r)=0,
    \qquad
    D''(r)\neq0.
    \label{eq:asym-forced-double-zero}
\end{equation}
Thus the dressed saddle in the central gap is forced by the orientation of
the physical spectral branch; it is not an additional condition imposed on
the solution.  We first use this saddle to construct the moment curve, then
determine its transverse geometry, and finally show that the Pick response
measures its conormal.

\subsection{Gap reality and the filling-fraction curve}
\label{subsec:asym-gap-reality}

Throughout the central gap the resolvent is real, and therefore
\begin{equation}
    D(x)=y(x)^2\geq0,
    \qquad x\in G.
    \label{eq:gap-reality}
\end{equation}
Solving this inequality for $m_2$ gives
\begin{equation}
    m_2\leq U_g(x;m_1),
    \label{eq:asym-gap-bound}
\end{equation}
where
\begin{equation}
    U_g(x;m_1)
    =\frac1g-m_1x
    +\frac{1-4g}{4g}x^2
    -\frac12x^4
    +\frac g4x^6.
    \label{eq:asym-U}
\end{equation}
Hence a two-cut solution with gap $G$ must satisfy
\begin{equation}
    m_2\leq\inf_{x\in G}U_g(x;m_1).
    \label{eq:asym-gap-inf}
\end{equation}
On the other hand, the forced zero $y(r)=0$ gives
\begin{equation}
    m_2=U_g(r;m_1)\geq\inf_{x\in G}U_g(x;m_1).
\end{equation}
The two inequalities therefore meet:
\begin{equation}
    m_2=\inf_{x\in G}U_g(x;m_1),
    \qquad
    \partial_xU_g(r;m_1)=0.
    \label{eq:asym-envelope-conditions}
\end{equation}
Thus saturation of the gap-reality envelope follows directly from the
global sign of the physical branch.

The stationarity condition gives
\begin{equation}
    m_1(r)
    =\frac{r}{2g}
    \left(3g^2r^4-4gr^2+1-4g\right),
    \label{eq:asym-m1r}
\end{equation}
and substitution into $m_2=U_g(r;m_1)$ gives
\begin{equation}
    m_2(r)
    =-\frac{5g^2r^6-6gr^4+(1-4g)r^2-4}{4g}.
    \label{eq:asym-m2r}
\end{equation}
Equivalently, the discriminant factorizes as
\begin{equation}
    D(z)=g^2(z-r)^2Q_4(z),
    \label{eq:asym-factorization}
\end{equation}
with
\begin{align}
    Q_4(z)={}&z^4+2rz^3+
    \left(3r^2-\frac2g\right)z^2
    +\left(4r^3-\frac{4r}{g}\right)z
    \notag\\
    &+5r^4-\frac{6r^2}{g}
    +\frac{1-4g}{g^2}.
    \label{eq:asym-Q4}
\end{align}
The curve \eqref{eq:asym-m1r}--\eqref{eq:asym-m2r} is the usual
one-parameter family of regular two-cut solutions and agrees with the analytic
cut/zero construction of Kazakov and Zheng \cite{KazakovZheng2022}. The
parameter $r$ is a convenient coordinate on this family, but is not itself the
filling fraction. The physical segment is the range for which $Q_4$ has four
ordered real roots, $r$ lies between the two cuts, and the density is
nonnegative.

The local convexity of the envelope is
\begin{equation}
    U_g''(r)=\frac g2Q_4(r).
    \label{eq:asym-convexity-Q}
\end{equation}
Thus $Q_4(r)>0$ for a regular saddle in a genuine gap. When
$Q_4(r)=0$, the dressed saddle collides with a branch point and the regular
geometry degenerates.

\subsection{Moment-space geometry and transverse analyticity}
\label{subsec:asym-analyticity}

The low-moment space is two-dimensional, with coordinates $(m_1,m_2)$. A
regular two-cut solution has one dressed saddle and therefore one transverse
condition, leaving the expected one-dimensional filling-fraction family.
The local conormal follows directly from the double-zero condition. Since
\begin{equation}
    \frac{\partial D}{\partial m_1}=-4gz,
    \qquad
    \frac{\partial D}{\partial m_2}=-4g,
\end{equation}
a variation tangent to the physical family satisfies
\begin{equation}
    r\,\delta m_1+\delta m_2=0.
    \label{eq:asym-normal-coordinate}
\end{equation}
Hence the conormal covector is
\begin{equation}
    \boldsymbol n(r)=g(r,1),
\end{equation}
and the slope of the filling-fraction curve is
\begin{equation}
    \frac{\dd m_2}{\dd m_1}=-r.
    \label{eq:asym-slope}
\end{equation}
The same relation follows by differentiating
$m_2=U_g(r;m_1)$ and using $\partial_rU_g=0$.

The two sides of this curve have a simple local interpretation. Perturbing
a physical point by $(\delta m_1,\delta m_2)$ gives, on the generic
transverse splitting scale $z-r=O(\|\delta\mathbf m\|^{1/2})$,
\begin{equation}
    D(z)
    =g^2Q_4(r)(z-r)^2
    -4g\bigl(r\,\delta m_1+\delta m_2\bigr)
    +O\!\left(\|\delta\mathbf m\|^{3/2}\right).
    \label{eq:asym-local-splitting}
\end{equation}
Writing
\begin{equation}
    \delta n=r\,\delta m_1+\delta m_2,
\end{equation}
the split zeros satisfy
\begin{equation}
    (z-r)^2\simeq\frac{4\delta n}{gQ_4(r)}.
    \label{eq:asym-root-splitting}
\end{equation}
For $\delta n>0$, $D(r)<0$ and the candidate no longer satisfies reality
at the old gap point. For $\delta n<0$, the double zero splits into a
complex-conjugate pair of simple zeros, one in the upper half-plane, which
would be a branch point of the resolvent and is therefore excluded by
analyticity. Thus
\begin{equation}
    \begin{array}{rcl}
    \delta n>0 &\Longrightarrow& \text{gap reality fails},\\[1mm]
    \delta n<0 &\Longrightarrow& \text{upper-half-plane analyticity fails},\\[1mm]
    \delta n=0 &\Longrightarrow& \text{tangent motion to first order}.
    \end{array}
    \label{eq:asym-two-sided-exclusion}
\end{equation}
Reality and analyticity therefore localize the physical family
transversely in moment space.

\subsection{Pick response and the conormal direction}
\label{subsec:asym-pick}

The Pick response is
\begin{equation}
    P(z,z)=-\frac{\operatorname{Im}W(z)}{\operatorname{Im}z}
    =\int_{\mathbb R}\frac{\rho(x)\,\dd x}{|z-x|^2}\geq0.
    \label{eq:asym-diagonal-pick}
\end{equation}
For the asymmetric quartic the natural probe approaches the real dressed
saddle from the upper half-plane,
\begin{equation}
    z=E+\ii\eta,
    \qquad E\simeq r.
\end{equation}
Using
\begin{equation}
    \partial_{m_a}W(z)
    =-\frac{1}{4y(z)}\partial_{m_a}D(z),
\end{equation}
and
\begin{equation}
    y(z)=a_r(z-r)+O\bigl((z-r)^2\bigr),
    \qquad a_r=y'(r)\neq0,
\end{equation}
we obtain the universal local profile
\begin{equation}
    \nabla_{(m_1,m_2)}P(E+\ii\eta,E+\ii\eta)
    =\frac{g}{a_r}\,
    \frac{(r,1)}{(E-r)^2+\eta^2}
    +O(1),
    \label{eq:asym-pick-lorentzian}
\end{equation}
for $E-r=O(\eta)$. In particular,
\begin{equation}
    \nabla_{(m_1,m_2)}P(r+\ii\eta,r+\ii\eta)
    =\frac{g}{a_r\eta^2}(r,1)+O(1).
    \label{eq:asym-pick-normal}
\end{equation}
The enhanced response is therefore aligned with the conormal of the
filling-fraction curve. Contracting with a tangent vector
\begin{equation}
    \boldsymbol t=(1,-r)
\end{equation}
removes the leading singularity,
\begin{equation}
    \boldsymbol n(r)\cdot\boldsymbol t=0.
\end{equation}
Thus the dressed saddle is a regular point of the resolvent but a singular
point of its transverse moment sensitivity. The Cauchy probe does not select
the filling fraction; it measures the conormal geometry of the family with a
resolution set by $\eta$.

\subsection{Filling fraction and endpoints of the physical family}
\label{subsec:asym-filling-fraction}

The constraints above determine the one-dimensional family
\begin{equation}
    \bigl(m_1(r),m_2(r)\bigr),
\end{equation}
but they do not select a particular value of $r$. This remaining direction
is the filling-fraction modulus. The equilibrium solution is selected by the
global period condition
\begin{equation}
    \Pi(r)=\int_{e_2}^{e_3}y(x)\,\dd x=0.
    \label{eq:asym-period}
\end{equation}
For the symmetric potential this gives the equal-filling point
\begin{equation}
    r=0,
    \qquad m_1=0,
    \qquad m_2=\frac1g.
\end{equation}

The physical family terminates when the spectral geometry degenerates. For
$1/15<g<1/4$, the dressed saddle reaches a branch point,
\begin{equation}
    Q_4(r)=0,
\end{equation}
or equivalently
\begin{equation}
    15g^2r^4-12gr^2+1-4g=0,
    \label{eq:asym-endpoint}
\end{equation}
with
\begin{equation}
    r_c^2=\frac{6-\sqrt{21+60g}}{15g}.
\end{equation}
At this point the zero is no longer a regular nonbranching dressed saddle.
For $0<g<1/15$, one of the cuts collapses first; this degeneration is detected
by the discriminant of $Q_4$ and is not localized at the dressed saddle.
These two endpoint mechanisms illustrate the complementary roles of local
Cauchy probes and the global root structure of the spectral curve.

\section{The sextic: multidimensional moment space}
\label{sec:sextic}

The asymmetric quartic has one tangent direction and one conormal direction.
To test the reconstruction in higher dimension, we now turn to a sextic
potential, for which the planar loop equation leaves more than one
independent low moment.  We first analyze its $\mathbb Z_2$-symmetric sector
and then restore all four low moments in a three-cut example.

We begin with the even potential
\begin{equation}
    V(x)
    =
    \frac{x^6}{6}
    +\frac{b}{4}x^4
    +\frac{\mu}{2}x^2,
    \qquad
    V'(z)=z^5+bz^3+\mu z,
    \label{eq:sextic-potential}
\end{equation}
and first restrict to the $\mathbb Z_2$-symmetric sector,
\begin{equation}
    m_{2k+1}=0.
\end{equation}
The two unfixed moments are then $(m_2,m_4)$.  Different dressed saddles
define different conormal directions in this moment plane.  The model also
shows why the full finite complex plane is needed: as the couplings vary,
the relevant saddles need not remain on a distinguished symmetry axis.

\subsection{Loop equation and the reality envelope}
\label{subsec:sextic-setup}

The loop polynomial is
\begin{equation}
    \mathcal Q(z)
    =
    z^4+(m_2+b)z^2
    +m_4+b m_2+\mu,
    \label{eq:sextic-loop-polynomial}
\end{equation}
and the discriminant is
\begin{align}
    D(z)
    ={}&
    z^{10}+2bz^8+(b^2+2\mu)z^6
    +(2b\mu-4)z^4
    \notag\\
    &+
    \left[\mu^2-4(m_2+b)\right]z^2
    -4(m_4+b m_2+\mu).
    \label{eq:sextic-D}
\end{align}
Because $D$ is even, it is convenient to introduce
\begin{equation}
    s=z^2
\end{equation}
and
\begin{align}
    \widetilde D(s)
    ={}&
    s^5+2bs^4+(b^2+2\mu)s^3
    +(2b\mu-4)s^2
    \notag\\
    &+
    \left[\mu^2-4(m_2+b)\right]s
    -4(m_4+b m_2+\mu).
    \label{eq:sextic-Dtilde}
\end{align}
The two moments enter only the last two coefficients and do so
affinely.

On the imaginary axis set
\begin{equation}
    q=\eta^2,
    \qquad
    \Delta(q)\equiv-D(\ii\sqrt q).
\end{equation}
Then
\begin{align}
    \Delta(q)
    ={}&
    F_{b,\mu}(q)
    +4\left[m_4+(b-q)m_2\right],
    \label{eq:sextic-Delta}
\end{align}
where
\begin{equation}
    F_{b,\mu}(q)
    =
    q^5-2bq^4+(b^2+2\mu)q^3
    +(4-2b\mu)q^2
    +(\mu^2-4b)q+4\mu .
    \label{eq:sextic-F}
\end{equation}
Since $y(\ii\eta)$ is purely imaginary,
\begin{equation}
    D(\ii\eta)\leq0,
\end{equation}
and every $q\geq0$ gives the half-plane constraint
\begin{equation}
    m_4+(b-q)m_2
    \geq
    -\frac14F_{b,\mu}(q).
    \label{eq:sextic-reality-halfplane}
\end{equation}
The admissible region obtained from imaginary-axis reality is therefore
\begin{equation}
    m_4
    \geq
    \sup_{q\geq0}
    \left[
        (q-b)m_2-\frac14F_{b,\mu}(q)
    \right].
    \label{eq:sextic-convex-envelope}
\end{equation}
At a regular interior tangency $q=q_\star>0$, with
$\Delta''(q_\star)\neq0$,
\begin{equation}
    \Delta(q_\star)=0,
    \qquad
    \Delta'(q_\star)=0,
    \label{eq:sextic-envelope-tangency}
\end{equation}
which is precisely the condition that $D$ have a nonbranching double
zero at $z=\pm\ii\sqrt{q_\star}$.

The boundary value $q=0$ must be treated separately because
$q=-z^2$ changes root multiplicities at the origin. Its reality constraint is
\begin{equation}
    m_4+b m_2+\mu\geq0.
    \label{eq:sextic-linear-bound}
\end{equation}

\subsection{One cut: two dressed-saddle constraints}
\label{subsec:sextic-onecut}

For a symmetric one-cut solution,
\begin{equation}
    \operatorname{supp}\rho
    =
    [-\sqrt{s_0},\sqrt{s_0}],
\end{equation}
there are two simple branch points.  Since the sextic has
\[
\deg M_{d-s-1}=4,
\]
the remaining zeros of the discriminant must be nonbranching.  In the
regular case this gives
\begin{equation}
    \widetilde D(s)
    =
    (s^2+cs+d)^2(s-s_0).
    \label{eq:sextic-onecut-factorization}
\end{equation}
Matching the coefficients gives
\begin{equation}
    c=b+\frac{s_0}{2},
    \qquad
    d=
    \mu+\frac{b s_0}{2}
    +\frac{3s_0^2}{8},
    \label{eq:sextic-cd}
\end{equation}
together with
\begin{equation}
    5s_0^3+6b s_0^2+8\mu s_0=32.
    \label{eq:sextic-onecut-normalization}
\end{equation}
The two moments are
\begin{equation}
    m_2
    =
    \frac{s_0}{256}
    \left(
       64+4b s_0^2+5s_0^3
    \right),
    \label{eq:sextic-m2-onecut}
\end{equation}
and
\begin{equation}
    m_4
    =
    \frac{s_0^2}{256}
    \left(
       32+3b s_0^2+4s_0^3
    \right).
    \label{eq:sextic-m4-onecut}
\end{equation}
The density is
\begin{equation}
    \rho(x)
    =
    \frac{1}{2\pi}
    (x^4+c x^2+d)
    \sqrt{s_0-x^2}.
    \label{eq:sextic-onecut-density}
\end{equation}
For $b>0$, positivity reduces to $d\geq0$; for general $b$ the
quadratic in $x^2$ must be nonnegative throughout the cut.

The two roots of
\begin{equation}
    s^2+cs+d=0
\end{equation}
are the two dressed-saddle values in the $s$-plane.  If they are
distinct, each supplies one independent constraint in the
$(m_2,m_4)$ plane.  Indeed,
\begin{equation}
    \partial_{m_2}D=-4(z^2+b),
    \qquad
    \partial_{m_4}D=-4,
\end{equation}
so a regular dressed saddle $z_a$ has conormal
\begin{equation}
    \boldsymbol n_a=(z_a^2+b,1).
    \label{eq:sextic-normal-vector}
\end{equation}
For two distinct values $s_a=z_a^2$,
\begin{equation}
    (s_1+b)\delta m_2+\delta m_4=0,
    \qquad
    (s_2+b)\delta m_2+\delta m_4=0,
\end{equation}
imply
\begin{equation}
    \delta m_2=\delta m_4=0.
    \label{eq:sextic-local-rigidity}
\end{equation}
Thus the two regular dressed saddles isolate the one-cut solution
locally in the two-dimensional moment space.

When the two roots of $s^2+cs+d$ are negative,
\begin{equation}
    s=-q_1,
    \qquad
    s=-q_2,
    \qquad
    q_1,q_2>0,
\end{equation}
the dressed saddles lie on the imaginary axis.  The physical branch may
then be written as
\begin{equation}
    y(\ii\eta)
    =
    \ii\sqrt{\eta^2+s_0}\,
    \left(
       \eta^4-c\eta^2+d
    \right).
    \label{eq:sextic-onecut-y-imag}
\end{equation}
Consequently $y(\ii\eta)$ changes sign at each $q_a=\eta^2$, and the two
analyticity conditions appear directly as two tangencies of the
reality envelope,
\begin{equation}
    \Delta(q_a)=\Delta'(q_a)=0,
    \qquad a=1,2.
    \label{eq:sextic-two-tangencies}
\end{equation}

For example, for
\begin{equation}
    b=5,
    \qquad
    s_0=1,
\end{equation}
one obtains
\begin{equation}
    \mu=-\frac38,
    \qquad
    m_2=\frac{89}{256},
    \qquad
    m_4=\frac{51}{256},
\end{equation}
and
\begin{equation}
    \Delta(q)
    =
    (q+1)(q-5)^2
    \left(q-\frac12\right)^2.
    \label{eq:sextic-simple-Delta}
\end{equation}
The two active heights are therefore $q=1/2$ and $q=5$, and their
independent tangencies fix the two moments.

The dressed saddles need not remain on the imaginary axis.  The roots
\begin{equation}
    s_\pm
    =
    \frac{-c\pm\sqrt{c^2-4d}}{2}
\end{equation}
form a complex-conjugate pair when $c^2-4d<0$.  This does not by itself
signal a change of cut topology: the physical density may remain
regular and positive.  In this regime the imaginary-axis envelope no
longer displays the two dressed-saddle conditions as tangencies, but
the zeros remain visible in the full finite-plane analytic structure.

We now restrict to $b>0$ and the central one-cut/two-cut transition.
Then $c=b+s_0/2>0$, so the prefactor in
\eqref{eq:sextic-onecut-density} first vanishes at the origin. The
one-cut phase reaches this boundary when
\begin{equation}
    d=0,
\end{equation}
or equivalently
\begin{equation}
    s_c^2(s_c+b)=16.
    \label{eq:sextic-critical-condition}
\end{equation}

\subsection{Two cuts and the critical merger}
\label{subsec:sextic-twocut}

Keeping $b>0$, sufficiently negative $\mu$ gives a symmetric two-cut solution,
\begin{equation}
    [-B,-A]\cup[A,B].
\end{equation}
Oddness of the physical spectral function implies
\begin{equation}
    y(0)=0,
\end{equation}
so the origin is a dressed saddle in the central gap.  The regular
two-cut factorization is
\begin{equation}
    \widetilde D(s)
    =
    s(s+C)^2(s-A^2)(s-B^2).
    \label{eq:sextic-twocut-factorization}
\end{equation}
The root at $s=0$ gives immediately
\begin{equation}
    m_4+b m_2+\mu=0.
    \label{eq:sextic-twocut-linear}
\end{equation}
Thus the $q=0$ reality condition is saturated throughout the two-cut
phase.

In this regime $C>0$, as also follows from the endpoint conditions
below. The remaining pair of dressed saddles is at
\begin{equation}
    z=\pm\ii\sqrt C.
\end{equation}
Its existence can also be seen directly from the sign of the physical
branch.  With the square root fixed by its large-$z$ behavior,
\begin{equation}
    y(\ii\eta)
    =
    -\ii\eta(C-\eta^2)
    \sqrt{(\eta^2+A^2)(\eta^2+B^2)}.
    \label{eq:sextic-twocut-y-imag}
\end{equation}
Hence
\begin{equation}
    \operatorname{Im}y(\ii\eta)<0
    \quad (0<\eta<\sqrt C),
\end{equation}
while
\begin{equation}
    \operatorname{Im}y(\ii\eta)>0
    \quad (\eta>\sqrt C).
\end{equation}
The physical branch therefore passes through
\begin{equation}
    y(\ii\sqrt C)=0,
\end{equation}
and the corresponding reality constraint is saturated:
\begin{equation}
    \Delta(C)=\Delta'(C)=0.
\end{equation}
The two independent dressed-saddle conditions are thus the real saddle
at $z=0$ and the conjugate pair at $z=\pm\ii\sqrt C$.

Matching coefficients gives
\begin{equation}
    2C^3-3bC^2+(b^2+2\mu)C-b\mu+2=0,
    \label{eq:sextic-C-cubic}
\end{equation}
or, equivalently,
\begin{equation}
    \mu(C)
    =
    -C(C-b)-\frac{2}{2C-b},
    \label{eq:sextic-mu-C}
\end{equation}
\begin{equation}
    m_2(C)
    =
    C-b+\frac{1}{(2C-b)^2},
    \label{eq:sextic-m2-C}
\end{equation}
and
\begin{equation}
    m_4(C)
    =
    -b\,m_2(C)-\mu(C).
    \label{eq:sextic-m4-C}
\end{equation}
The endpoints are
\begin{equation}
    A^2
    =
    C-b-\frac{2}{\sqrt{2C-b}},
    \qquad
    B^2
    =
    C-b+\frac{2}{\sqrt{2C-b}}.
    \label{eq:sextic-AB-C}
\end{equation}
The physical branch requires
\begin{equation}
    2C-b>0,
    \qquad
    A^2>0,
    \qquad
    B^2>A^2,
\end{equation}
together with positivity of the density.

The two-cut counting is therefore particularly transparent:
\begin{equation}
    z=0
    \quad+\quad
    z=\pm\ii\sqrt C
    \quad\Longrightarrow\quad
    (m_2,m_4).
\end{equation}
The two corresponding conormals,
\begin{equation}
    (b,1),
    \qquad
    (b-C,1),
\end{equation}
are independent for $C\neq0$.

The one- and two-cut branches meet when
\begin{equation}
    A\rightarrow0.
\end{equation}
Using \eqref{eq:sextic-AB-C}, this gives precisely
\begin{equation}
    s_c^2(s_c+b)=16,
\end{equation}
the same condition found from $d=0$ on the one-cut side.  At the
transition the central dressed saddle collides with the two inner
branch points, and the discriminant develops a fourth-order zero at
the origin,
\begin{equation}
    D(z)\sim z^4.
\end{equation}
For $b=5$,
\begin{equation}
    s_c=\frac{\sqrt{17}-1}{2},
    \qquad
    \mu_c=-\frac{7+17\sqrt{17}}{16}.
\end{equation}

\subsection{Pick response and dressed-saddle geometry}
\label{subsec:sextic-pick}

The dependence of the diagonal response on the two low moments is
\begin{equation}
    \frac{\partial P}{\partial m_2}
    =
    -\frac1{\eta}
    \operatorname{Im}\left[
       \frac{z^2+b}{y(z)}
    \right],
    \qquad
    \frac{\partial P}{\partial m_4}
    =
    -\frac1{\eta}
    \operatorname{Im}\left[
       \frac1{y(z)}
    \right].
    \label{eq:sextic-pick-sensitivity}
\end{equation}
Thus the same covector
\begin{equation}
    \boldsymbol n(z_a)=(z_a^2+b,1)
\end{equation}
that defines the transverse constraint also determines the direction
of enhanced Cauchy sensitivity.

For a real regular dressed saddle $r$, approached as
\begin{equation}
    z=E+\ii\eta,
    \qquad
    E-r=O(\eta),
\end{equation}
one has
\begin{equation}
    y(z)=a_r(z-r)+O((z-r)^2),
\end{equation}
and therefore
\begin{equation}
    \nabla_{(m_2,m_4)}
    P(E+\ii\eta,E+\ii\eta)
    =
    \frac{1}{a_r}
    \frac{(r^2+b,1)}
         {(E-r)^2+\eta^2}
    +O(1).
    \label{eq:sextic-real-saddle-response}
\end{equation}
The response is Lorentzian, with height $O(\eta^{-2})$, and is aligned
with the conormal to the physical moment locus.

For a nonreal dressed saddle $z_a$ in the upper half-plane,
\begin{equation}
    \nabla_{\mathbf m}W(z)
    \sim
    \frac{\boldsymbol n(z_a)}{a_a(z-z_a)},
\end{equation}
so the moment susceptibility of $P$ is singular as the probe approaches
$z_a$,
\begin{equation}
    \|\nabla_{\mathbf m}P(z,z)\|
    =
    O(|z-z_a|^{-1})
\end{equation}
generically.  The finite-plane probe therefore continues to detect the
dressed saddles after they have migrated away from the imaginary axis.
What is lost in that regime is only their representation as tangencies
of the one-dimensional imaginary-axis envelope.

\subsection{Three cuts and the filling-fraction manifold}
\label{subsec:sextic-three-cut}

The role of the dressed-saddle constraints becomes particularly clear
in a three-cut sextic. We now leave the $b>0$ restriction, allow the
fillings to vary, and restore
the four low moments
\begin{equation}
    \mathbf m=(m_1,m_2,m_3,m_4).
\end{equation}
For convenience, take
\begin{equation}
    V'(z)
    =
    \lambda
    \left(
        z^5+bz^3+\mu z+h
    \right),
    \qquad
    \lambda>0,
    \label{eq:threecut-potential}
\end{equation}
with loop polynomial
\begin{align}
    \mathcal Q(z)
    =
    \lambda\Big[
        &z^4+m_1z^3+(m_2+b)z^2
        \notag\\
        &+(m_3+b m_1)z
        +(m_4+b m_2+\mu)
    \Big].
    \label{eq:threecut-loop-polynomial}
\end{align}

Let the support consist of three ordered cuts,
\begin{equation}
    [e_1,e_2]\cup[e_3,e_4]\cup[e_5,e_6].
\end{equation}
There are two real gaps,
\begin{equation}
    G_1=(e_2,e_3),
    \qquad
    G_2=(e_4,e_5).
\end{equation}
Positivity of the density fixes the orientation of the physical
spectral branch at the endpoints.  In each gap,
\begin{equation}
    y(e_{2i}+\epsilon)>0,
    \qquad
    y(e_{2i+1}-\epsilon)<0,
    \qquad
    i=1,2.
\end{equation}
For sufficiently small $\epsilon>0$, continuity therefore forces a zero
\begin{equation}
    r_i\in G_i
\end{equation}
in each gap.  Since a regular three-cut sextic has
\begin{equation}
    \deg M_{d-s-1}=2,
\end{equation}
these are precisely the two dressed saddles:
\begin{equation}
    D(r_i)=D'(r_i)=0,
    \qquad
    i=1,2.
    \label{eq:threecut-factorization}
\end{equation}

The moment dependence of $D$ is controlled by the covector
\begin{equation}
    \boldsymbol n(r)
    =
    \lambda\Big(
       r(r^2+b),\,
       r^2+b,\,
       r,\,
       1
    \Big).
    \label{eq:threecut-normal}
\end{equation}
A tangent variation of a fixed three-cut solution therefore obeys
\begin{equation}
    \boldsymbol n(r_i)\cdot\delta\mathbf m=0,
    \qquad
    i=1,2.
    \label{eq:threecut-transverse-conditions}
\end{equation}
Collecting the two conormals,
\begin{equation}
    N(r_1,r_2)
    =
    \lambda\begin{pmatrix}
    r_1(r_1^2+b) & r_1^2+b & r_1 & 1\\
    r_2(r_2^2+b) & r_2^2+b & r_2 & 1
    \end{pmatrix},
\end{equation}
one has
\begin{equation}
    \operatorname{rank}N=2
\end{equation}
for $r_1\neq r_2$.  Consequently,
\begin{equation}
    T_{\mathbf m}\mathcal F_3
    =
    \ker N,
    \qquad
    \dim\mathcal F_3=4-2=2.
    \label{eq:threecut-tangent-space}
\end{equation}
The two remaining directions are precisely the two filling-fraction
moduli.

Cauchy probes approaching the two real dressed saddles recover the
same conormal geometry.  Writing
\begin{equation}
    y(z)=a_i(z-r_i)+\cdots,
\end{equation}
one finds
\begin{equation}
    \nabla_{\mathbf m}
    P(E+\ii\eta,E+\ii\eta)
    =
    \frac{1}{a_i}
    \frac{\boldsymbol n(r_i)}
         {(E-r_i)^2+\eta^2}
    +O(1),
    \qquad
    E-r_i=O(\eta).
    \label{eq:threecut-Pick-normal}
\end{equation}
The two susceptibility peaks therefore span the two-dimensional
conormal space, while their common null space is the tangent space of
the filling-fraction manifold.

The local dressed-saddle conditions do not determine the two independent
filling fractions.  Unconstrained equilibrium requires the effective
potential to take the same constant value on all three occupied cuts.  This
gives the two independent period conditions
\begin{equation}
    \Pi_1
    =
    \int_{e_2}^{e_3}y(x)\,\dd x=0,
    \qquad
    \Pi_2
    =
    \int_{e_4}^{e_5}y(x)\,\dd x=0.
    \label{eq:threecut-periods}
\end{equation}
These conditions fix the two filling-fraction moduli and isolate the
equal-chemical-potential point on $\mathcal F_3$.  This point represents the
global equilibrium measure provided that the associated effective potential
also satisfies the variational inequality
$V_{\mathrm{eff}}(x)\geq\ell$ outside the support.

A useful exact example is obtained by choosing
\begin{equation}
    b=-\frac{4q^2}{3},
    \qquad
    \mu=\frac{q^4}{3},
    \qquad
    h=0.
\end{equation}
Defining
\begin{equation}
    w(z)=z^3-q^2z,
\end{equation}
the potential becomes
\begin{equation}
    V(z)=\frac{\lambda}{6}w(z)^2.
\end{equation}
At equilibrium,
\begin{equation}
    m_1=m_3=0,
    \qquad
    m_2=\frac{2q^2}{3},
    \qquad
    m_4=\frac{2q^4}{3},
\end{equation}
and the discriminant factorizes as
\begin{equation}
    \frac{D(z)}{\lambda^2}
    =
    \left(z^2-\frac{q^2}{3}\right)^2
    \left[
       z^2(z^2-q^2)^2-\tau
    \right],
    \qquad
    \tau=\frac4\lambda .
    \label{eq:threecut-exact-factorization}
\end{equation}
The two dressed saddles are
\begin{equation}
    r_\pm=\pm\frac{q}{\sqrt3}.
\end{equation}
The regular three-cut interpretation requires
\begin{equation}
    \tau<\frac{4q^6}{27}.
\end{equation}
At equality neighboring cuts merge and the dressed saddles become
degenerate. In the regular regime each of the three monotone
branches of $w(x)$ maps one cut onto
$[-\sqrt{\tau},\sqrt{\tau}]$. The density is
\begin{equation}
    \rho(x)=\frac{\lambda}{6\pi}|w'(x)|
       \sqrt{\tau-w(x)^2}.
    \label{eq:threecut-pullback-density}
\end{equation}
Changing variables separately on each cut therefore gives
\begin{equation}
    \varepsilon_j=\frac{\lambda}{6\pi}
       \int_{-\sqrt{\tau}}^{\sqrt{\tau}}\sqrt{\tau-w^2}\,\dd w
       =\frac{\lambda\tau}{12}=\frac13.
\end{equation}
Thus the cubic pullback, rather than reflection symmetry alone, fixes
\begin{equation}
    \varepsilon_1=\varepsilon_2=\varepsilon_3=\frac13.
\end{equation}

\section{Conclusion}
\label{sec:conclusion}

The cut and zero structure of a planar one-matrix resolvent is classically
encoded in the multiplicities and ordering of the zeros of its discriminant.
In particular, using the conditions $D(r)=D'(r)=0$ to constrain the low
moments is standard and appears explicitly in the analytic bootstrap of
Kazakov and Zheng~\cite{KazakovZheng2022}.  The result developed here is a
differential refinement of that picture.  On a regular fixed-topology
family, the double-zero constraints define conormal covectors in low-moment
space.  A Cauchy probe approaching a real dressed saddle has a universal
$\eta^{-2}$ response aligned with the corresponding conormal, whereas this
leading enhancement cancels along tangent filling-fraction deformations.
When the sampled real saddles supply a complete conormal span, the
row space of the rescaled response matrix reconstructs that space and its
kernel reconstructs the tangent space.

This distinction also fixes the role of the rational probe.  It is not a
positivity principle stronger than the Hankel hierarchy: after truncation to
the same polynomial space, the Cauchy and monomial Gram matrices are related
by a congruence transformation and define the same cone.  The advantage of
the exact Cauchy kernel is instead its finite-plane localization. The real
part of the probe coordinate $z$ selects a spectral region, its imaginary
part sets a continuous resolution scale, and moving the probe toward a
dressed saddle converts the local
spectral singularity of the moment derivative into a measurable transverse
direction.  The finite-node bootstrap of Appendix~\ref{appendixA} illustrates
both this selectivity and the associated numerical costs of branch tracking,
node design, and conditioning.

The examples separate the ingredients of the construction.  In the
symmetric quartic, the physical moments are already known, so the analysis
serves as a calibration: reality produces an envelope, the Pick sign and
continuation orient the physical sheet, and the distinguished dressed saddle
migrates between the imaginary axis, a real gap, and the exterior of a cut.
The asymmetric quartic is the first nontrivial geometric application.  Its
gap saddle parametrizes a one-dimensional two-cut family in the
$(m_1,m_2)$ plane; gap reality and upper-half-plane analyticity exclude the
two opposite transverse displacements, while the Cauchy susceptibility
measures the surviving conormal direction.  The filling fraction remains a
longitudinal modulus and is fixed only by the equilibrium period condition.

The sextic models show that the construction is not restricted to a
one-dimensional moment plane.  In the symmetric sector, distinct dressed
saddles give independent covectors in $(m_2,m_4)$.  In the general three-cut
example, four low moments are reduced by two independent gap saddles to a
two-dimensional filling-fraction manifold.  Cauchy probes at the two saddles
recover its two-dimensional conormal space, while two global period
conditions select the equilibrium point.  Degenerate saddles mark the
critical locus at which this regular counting ceases to apply.

The resulting organization is therefore hierarchical.  The loop equation
defines the ambient algebraic moment space; cut factorization and analyticity
determine its regular fixed-topology locus; reality and positivity select
admissible branches; exact Cauchy probes measure the local transverse
geometry; and period matching, together with the equilibrium
variational inequality, completes the selection of the filling fractions.  In a
multicut problem the natural intermediate object is consequently not a
single moment vector, but the filling-fraction manifold together with its
tangent and conormal spaces.  Dressed saddles provide the local bridge
between this moment-space geometry and finite-plane measurements of the
resolvent.

\paragraph{Data and code availability.}
The numerical data and the Python and Wolfram Language codes used to generate
the results of Appendix~\ref{appendixA} are available on Zenodo
\cite{NassarPickData2026}. No external datasets were used.

\section*{Acknowledgments}

I would like to acknowledge support through the ICTP--Arab Fund Associates Programme (ARF01--AFESD Grant No. 14/2023) in the context of the project ``Advancing the Capabilities of Arab Researchers and Students''. I am grateful to Agnese Bissi for useful discussions and for valuable comments on the manuscript.

The author acknowledges the use of ChatGPT (OpenAI) for assistance with manuscript preparation and for the verification of numerical computations and figures. The author takes full responsibility for the scientific content, numerical results, and conclusions presented in the manuscript.

\appendix
\setcounter{figure}{0}
\setcounter{table}{0}

\renewcommand{\thefigure}{\Alph{section}.\arabic{figure}}
\renewcommand{\thetable}{\Alph{section}.\arabic{table}}
\renewcommand{\theHfigure}{\Alph{section}.\arabic{figure}}
\renewcommand{\theHtable}{\Alph{section}.\arabic{table}}

\section{Finite-node Pick bootstrap: effectiveness and numerical limitations}\label{appendixA}

This appendix supports the assessment in
Section~\ref{subsec:information-matched} with an explicit numerical test.
The Cauchy--Pick construction may be used not only as a local probe of
spectral geometry, but also as a direct bootstrap constraint on the
undetermined low moments.  To examine the effectiveness of this approach, we
apply it to the asymmetric quartic model
\begin{equation}
    V(x)=-\frac{x^2}{2}+\frac{g x^4}{4},
    \qquad g=0.1,
\end{equation}
without imposing the $\mathbb Z_2$ symmetry.  The planar loop equation leaves
$m_1$ and $m_2$ undetermined and gives
\begin{equation}
    W(z;m_1,m_2)
    =
    \frac12\left[gz^3-z-\sqrt{D(z;m_1,m_2)}\right],
    \label{eq:asym-pick-resolvent}
\end{equation}
where
\begin{equation}
    D(z;m_1,m_2)
    =
    g^2z^6-2gz^4+(1-4g)z^2
    -4gm_1z+4(1-gm_2).
    \label{eq:asym-pick-discriminant}
\end{equation}
For a collection of nodes $z_i\in\mathbb H_+$, we construct the Pick matrix
\begin{equation}
    P_{ij}(m_1,m_2)
    =
    \frac{\overline{W(z_i;m_1,m_2)}
          -W(z_j;m_1,m_2)}
         {z_j-\overline{z_i}}.
    \label{eq:asym-pick-matrix}
\end{equation}
For a physical positive spectral measure this matrix is a Gram matrix,
\begin{equation}
    P_{ij}
    =
    \int_{\mathbb R}
    \frac{\rho(x)\,\dd x}
         {(\overline z_i-x)(z_j-x)},
\end{equation}
and must therefore satisfy
\begin{equation}
    P(m_1,m_2)\succeq0.
    \label{eq:finite-pick-bootstrap-condition}
\end{equation}
A candidate with a nonfinite or nonpositive diagonal entry is rejected
before normalization. Otherwise, the trial pair $(m_1,m_2)$ is retained
when the smallest eigenvalue of the
diagonally normalized matrix
\begin{equation}
    \widehat P
    =
    \operatorname{diag}(P)^{-1/2}
    P\,
    \operatorname{diag}(P)^{-1/2},
    \label{eq:normalized-pick-matrix}
\end{equation}
obeys
\begin{equation}
    \lambda_{\min}(\widehat P)\geq-\tau_{\mathrm{PSD}},
    \qquad
    \tau_{\mathrm{PSD}}=10^{-8}.
\end{equation}
The diagonal normalization is essential:
without it, the eigenvalue test can be dominated by the widely different norms
of the Cauchy kernels associated with nodes at different heights.

\paragraph{Continuation of the physical sheet.}

The square root in \eqref{eq:asym-pick-resolvent} cannot be evaluated by making
an independent principal-branch choice at each node.  Instead, the physical
root is selected by its normalization
\begin{equation}
    W(z)\sim\frac{1}{z},
    \qquad |z|\longrightarrow\infty,
\end{equation}
and continued to every node along a radial path in the upper half-plane.  More
precisely, for a target node $z_i$ we discretize
\begin{equation}
    z(t)=t z_i,
    \qquad
    t_{\max}\geq t\geq1,
    \qquad
    |t_{\max}z_i|=60.
\end{equation}
The $t$ values are geometrically spaced between $t_{\max}$ and $1$;
the step counts below denote the number of sampled points on this path.
At the initial point, of the two roots
\begin{equation}
    W_\pm(z)
    =
    \frac12\left[V'(z)\pm\sqrt{D(z)}\right],
\end{equation}
we choose the one closest to $1/z$.  At every subsequent point, we choose the
root closest to the value selected at the preceding step.  This continuation
is performed independently for every node and every trial point in
$(m_1,m_2)$ space.  Repeating the calculation with $45$, $90$, and $180$
continuation steps gives no detectable difference between the $90$- and
$180$-step results on the sampled exact two-cut curve at double
precision. The accepted-point classifications on the full grid below are
also unchanged under this refinement. This checks the chosen numerical
paths but does not certify global analyticity of every accepted trial resolvent.

As an analytic benchmark, the regular two-cut family is obtained by requiring
a real nonbranching double zero $r$ of the discriminant,
\begin{equation}
    D(r;m_1,m_2)=0,
    \qquad
    \partial_zD(r;m_1,m_2)=0.
\end{equation}
Solving these equations gives the parametric moment curve
\begin{align}
    m_1(r)
    &=
    \frac{r}{2g}
    \left(3g^2r^4-4gr^2+1-4g\right),
    \label{eq:asym-exact-m1}
    \\
    m_2(r)
    &=
    -\frac{1}{4g}
    \left[
       5g^2r^6-6gr^4+(1-4g)r^2-4
    \right],
    \label{eq:asym-exact-m2}
\end{align}
with
\begin{equation}
    |r|<r_c,
    \qquad
    r_c=
    \sqrt{\frac{6-\sqrt{21+60g}}{15g}}.
    \label{eq:asym-physical-r-range}
\end{equation}
The closure includes the critical endpoints $|r|=r_c$. This exact
curve is not supplied to the numerical bootstrap; it is
superimposed afterward as an independent reference.

\paragraph{Contraction of the allowed region.}

\begin{figure}[t]
    \centering
    \includegraphics[width=\textwidth]{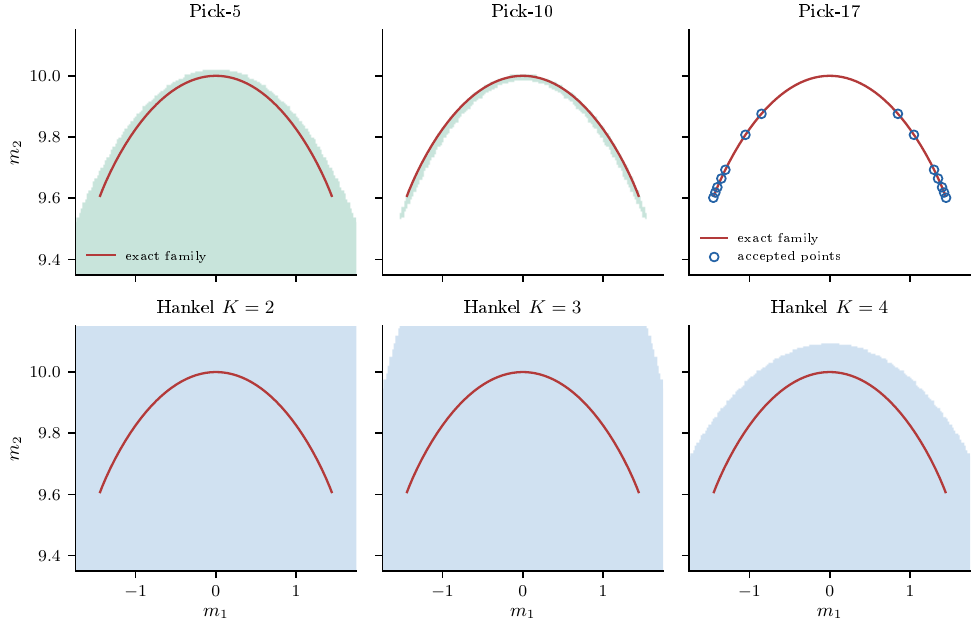}
    \caption{Finite Pick and Hankel tests for the asymmetric quartic at
    $g=0.1$. Shading marks accepted grid points for the indicated test.
    For Pick-17, the $14$ accepted points are shown by blue circles.
    The red curve is the independent analytic two-cut family obtained
    from $D(r)=D'(r)=0$. The comparison is not information matched:
    the Pick matrices use the exact algebraic resolvent, while each
    Hankel cutoff uses finitely many moments.}
    \label{fig:pick-hankel-asymmetric}
\end{figure}

We scan a $141\times141$ grid over
\begin{equation}
    -1.75\leq m_1\leq1.75,
    \qquad
    9.35\leq m_2\leq10.15.
\end{equation}
For the Hankel comparison, higher moments are generated using
\eqref{eq:planar-moment-recursion}; the Hankel matrices are diagonally
normalized and tested at the same threshold $-10^{-8}$.
Three nested sets containing $5$, $10$, and $17$ upper-half-plane nodes are
used:
\begin{align}
    \mathcal Z_5
    &=
    \{-4,-2,0,2,4\}+\ii,
    \\
    \mathcal Z_{10}
    &=
    \mathcal Z_5
    \cup
    \bigl(\{-3,-1.5,0,1.5,3\}+0.55\ii\bigr),
    \\
    \mathcal Z_{17}
    &=
    \mathcal Z_{10}
    \cup
    \bigl(\{-3,-2,-1,0,1,2,3\}+0.25\ii\bigr).
    \label{eq:pick-node-families}
\end{align}
The shifted ten-node comparison uses
\begin{equation}
    \begin{split}
    \mathcal Z_{10}^{\mathrm{sh}}
    ={}&\bigl(\{-3.6,-1.8,0.2,2.2,4.2\}+0.8\ii\bigr)\\
       &{}\cup\bigl(\{-2.7,-1.2,0.3,1.8,3.3\}+0.40\ii\bigr).
    \end{split}
\end{equation}
The resulting feasible regions are shown in
Fig.~\ref{fig:pick-hankel-asymmetric}.  The five-node condition retains a broad
two-dimensional region.  With ten nodes, this region contracts to a narrow
band that follows the exact two-cut moment curve.  The seventeen-node test accepts only $14$ grid points at the stated
tolerance; these are displayed as discrete markers, separately from the
continuous analytic reference curve.

The numerical results are summarized in
Table~\ref{tab:asym-pick-bootstrap}.  The feasible fraction is the fraction of
the chosen rectangular grid that survives the indicated test; it is therefore
a resolution- and window-dependent diagnostic rather than an intrinsic
measure of bootstrap strength.

\begin{table}[t]
    \centering
    \begin{tabular}{lccc}
        \hline
        Test
        & Feasible fraction
        & $\min_{\mathcal S_g}\lambda_{\min}(\widehat P)$
        & $\max_{\mathcal S_g}\kappa(\widehat P)$
        \\
        \hline
        Pick, $5$ nodes
        & $6.294955\times10^{-1}$
        & $4.0001\times10^{-3}$
        & $8.47\times10^{2}$
        \\
        Pick, $10$ nodes
        & $3.646698\times10^{-2}$
        & $4.0076\times10^{-7}$
        & $1.66\times10^{7}$
        \\
        Pick, $17$ nodes
        & $7.0419\times10^{-4}$
        & $1.8130\times10^{-12}$
        & $5.88\times10^{12}$
        \\
        Shifted Pick, $10$ nodes
        & $3.359992\times10^{-2}$
        & $1.2273\times10^{-6}$
        & $5.13\times10^{6}$
        \\
        Hankel, $K=4$
        & $7.702832\times10^{-1}$
        & --- & ---
        \\
        \hline
    \end{tabular}
    \caption{Numerical finite-node Pick bootstrap for the asymmetric quartic.
    Here $\mathcal S_g$ consists of $41$ equally spaced saddle values
    $r\in[-0.95r_c,0.95r_c]$ mapped to the exact two-cut moment curve.
    The eigenvalue and condition-number extrema are taken over this
    sample, not optimized over the continuous curve.
    $\kappa(\widehat P)$ is the spectral condition number of the normalized
    Pick matrix.  The feasible fractions refer specifically to the search
    window and grid specified in the text.}
    \label{tab:asym-pick-bootstrap}
\end{table}

The ten-node result provides a narrow band at substantially more moderate
conditioning than the seventeen-node test.  It retains only approximately
$3.65\%$ of the search grid and produces a thin band aligned with the exact
two-cut locus.  A shifted set of ten nodes retains approximately $3.36\%$.
Although the detailed boundaries depend on node placement, both designs
identify the same underlying curve.  This provides a nontrivial check that
the contraction is not an artifact of one specially tuned node configuration.

\paragraph{Advantages of the Pick bootstrap.}

The numerical experiment exhibits several useful features of finite-plane
positivity.

First, each Pick constraint uses the exact algebraic resolvent rather than a
finite truncation of its expansion at infinity.  Every matrix element
therefore resums information from the entire sequence of moments generated by
the loop equation.  This explains why a modest ten-node matrix can be much
more selective than a low-order Hankel test.

Second, the nodes are movable spectral probes.  Their real parts can be placed
near suspected gaps, saddles, or merging points, while their imaginary parts
set the resolution scale.  The bootstrap can consequently be adapted to the
spectral geometry under investigation instead of probing the measure only
through increasingly high monomials.

Third, the method applies directly to asymmetric solutions.  No parity
reduction is required, and the contraction occurs in the full
$(m_1,m_2)$ plane.  In the present example the Pick condition localizes the accepted grid
points near the one-dimensional family without imposing the double-zero
equations used to generate the analytic red curve.

Finally, Cauchy kernels can be better localized than high-degree monomials.
A carefully chosen finite node set may therefore detect local spectral
features with a comparatively small matrix.  This is precisely the property
that makes the same kernels effective as susceptibility probes of dressed
saddles.

\paragraph{Limitations and conditioning.}

These advantages come with additional numerical requirements. Through
\eqref{eq:asym-pick-resolvent}, the exact Pick matrix depends nonlinearly
on $(m_1,m_2)$ and requires a consistent algebraic sheet at every trial
point. The Hankel matrices also become nonlinear in these low moments
after loop-equation elimination, as explained in
Section~\ref{subsec:information-matched}. The additional Pick difficulty
is branch evaluation: pointwise principal square roots can jump between
sheets as the zeros of $D$ move, producing discontinuous constraints or
spurious positivity failures.

The matrices also become rapidly ill-conditioned as more nodes are added.
Along the exact two-cut family, the maximum condition number increases from
approximately $8.5\times10^2$ for five nodes to
$1.7\times10^7$ for ten nodes and $5.9\times10^{12}$ for seventeen nodes.
This deterioration arises because a large collection of nearby Cauchy
kernels becomes nearly linearly dependent.  Correspondingly, the smallest
positive eigenvalue on the exact curve falls from approximately
$4.0\times10^{-3}$ to $4.0\times10^{-7}$ and finally to
$1.8\times10^{-12}$. The seventeen-node matrix is therefore severely
ill-conditioned, but its condition number alone does not identify a
failure threshold of double precision. A $65$-digit evaluation at the
exact symmetric point $r=0$ gives
$\lambda_{\min}(\widehat P_{17})=1.8129692437513657\ldots\times10^{-12}$,
confirming a genuinely small positive eigenvalue. This pointwise check
does not certify the full trial grid.

The accepted-point count is sensitive to the eigenvalue threshold:
$74$, $14$, and $8$ points survive at thresholds $-10^{-6}$,
$-10^{-8}$, and $-10^{-10}$, respectively. The apparent contraction
must therefore be accompanied by tolerance and precision checks; the
smallest positive eigenvalue does not set the accuracy with which the
moment curve has been localized.

Conditioning also depends on the relative node positions, their distances
from the support, and the normalization of the kernels. Reducing node
heights improves spatial resolution, but its effect on conditioning is
not monotonic. For example, kernels whose nodes all lie far from a compact
support can become nearly proportional. Effective node selection must
therefore monitor both selectivity and the normalized Gram spectrum.

The comparison with the Hankel hierarchy also requires care.  A finite
Hankel matrix
\begin{equation}
    H^{(K)}_{ij}=m_{i+j},
    \qquad
    0\leq i,j\leq K,
\end{equation}
provides an affine semidefinite constraint in its full moment variables;
this does not make the low-moment feasible set convex after nonlinear
loop-equation elimination. The finite Pick calculation performed here evaluates
the exact, resummed algebraic resolvent.  The two calculations are therefore
not information matched.  The narrower ten-node Pick region does not imply
that Pick positivity is intrinsically stronger than moment positivity.
Indeed, when all nodes or all polynomial degrees are included, both
hierarchies express positivity of the same underlying spectral measure.  The
difference lies in how information is packaged at finite resolution.

\paragraph{Assessment.}

The asymmetric-quartic experiment shows that finite Pick matrices can provide
an effective bootstrap of the low moments: a moderate number of nodes
contracts a broad search region to a narrow neighborhood of the exact
two-cut family.  Their principal strengths are the use of the resummed
resolvent, flexible placement of spectral probes, and direct applicability to
asymmetric and multicut geometries.  Their principal weaknesses are nonlinear
dependence on the trial moments, the need for global sheet continuation,
sensitivity to node placement, and rapid loss of conditioning as the node set
is enlarged.

For systematic bounds, the polynomial Hankel formulation avoids algebraic
sheet tracking, while its loop equations must still be treated appropriately.
The present scan does not compare optimized or relaxed Hankel algorithms
at matched information or computational cost.  Finite Pick matrices are most
useful as a complementary tool: moderate, well-conditioned node sets can
sharpen a moment-space search and target specific regions of the spectral
plane, while the associated Cauchy susceptibilities provide local geometric
information that is not naturally exposed by a monomial basis.

\section{Detailed symmetric-quartic analysis}
\label{app:symmetric-quartic-details}

This appendix contains the algebraic steps underlying the compact
calibration in Section~\ref{sec:quartic-diagonal-pick}.  The details are
retained here to make the envelope optimization, branch orientation, and
root migration fully reproducible without interrupting the progression to
the asymmetric and multidimensional examples.

Consider the quartic potential
\begin{equation}
 V(x)=\frac{\mu}{2}x^2+\frac{g}{4}x^4,
 \qquad \mu=\pm1.
 \label{eq:quartic-general-potential}
\end{equation}
The choice $\mu=+1$ gives the single-well branch.  We first take $g>0$
and later follow its regular planar continuation into
$-1/12<g<0$.  The choice $\mu=-1$ gives the double-well model, for
which we take $g>0$.  We restrict throughout this
section to the $\mathbb{Z}_2$-symmetric planar branch,
\begin{equation}
 \rho(-x)=\rho(x),
 \qquad
 m_{2k+1}=0,
 \qquad
 m_n=\int_{\mathbb{R}}x^n\rho(x)\,\dd x,
 \qquad
 m_0=1.
 \label{eq:quartic-even-branch}
\end{equation}

\subsection{Reality on the imaginary axis and the lower envelope}

The planar loop equation is
\begin{equation}
 W(z)^2-V'(z)W(z)+ \mathcal{Q}(z)=0,
 \label{eq:quartic-planar-loop-equation}
\end{equation}
where
\begin{align}
 \mathcal{Q}(z)
 &=\int_{\mathbb{R}}\rho(x)
 \frac{V'(z)-V'(x)}{z-x}\,\dd x \notag\\
 &=\mu+g\bigl(z^2+z m_1+m_2\bigr).
\end{align}
Since $m_1=0$ on the symmetric branch, we obtain
\begin{equation}
 W(z)^2-\bigl(\mu z+gz^3\bigr)W(z)
 +\mu+g\bigl(z^2+m_2\bigr)=0.
 \label{eq:quartic-symmetric-loop-equation}
\end{equation}
With $y(z)=V'(z)-2W(z)$, the spectral curve is $y(z)^2=\Delta_{\mu,g}(z)$, with
\begin{align}
 \Delta_{\mu,g}(z)
 &=\bigl(\mu z+gz^3\bigr)^2
 -4\left[\mu+g\bigl(z^2+m_2\bigr)\right] \notag\\
 &=g^2z^6+2\mu g z^4+(1-4g)z^2-4\mu-4gm_2.
 \label{eq:quartic-general-discriminant}
\end{align}
The physical sheet is fixed by
\begin{equation}
 W(z)=\frac{1}{z}+O(z^{-3}),
 \qquad
 y(z)=V'(z)-\frac{2}{z}+O(z^{-3}),
 \qquad z\to\infty.
 \label{eq:quartic-physical-sheet}
\end{equation}

Two elementary properties of $W$ will be used throughout. Since $\rho$ is a
real measure on the real axis,
\begin{equation}
  W(\overline{z})=\overline{W(z)},
  \label{eq:reality}
\end{equation}
which holds for every such real spectral measure. If in addition the density is even,
$\rho(-x)=\rho(x)$, then substituting $x\to-x$ in
$W(z)=\int\rho(x)\,\dd x/(z-x)$ gives
\begin{equation}
  W(-z)=-W(z).
  \label{eq:oddness}
\end{equation}
Combining \eqref{eq:reality} and \eqref{eq:oddness} at $z=\ii\eta$,
\begin{equation}
  \overline{W(\ii\eta)}=W(-\ii\eta)=-W(\ii\eta)
  \qquad\Longrightarrow\qquad
  W(\ii\eta)\in\ii\mathbb{R}.
  \label{eq:quartic-imaginary-resolvent}
\end{equation}

We stress that \eqref{eq:oddness} is a statement about the \emph{solution},
not about the potential. In the quartic double-well two-cut regime, an even $V$ admits a
family of planar solutions with varying fillings. Only its
$\mathbb{Z}_2$-symmetric member, with
$\varepsilon_1=\varepsilon_2$, has an even density.  For
asymmetric members we have $m_1\neq0$, so $W(-z)\neq-W(z)$ and
\eqref{eq:quartic-imaginary-resolvent} fails: $W(\ii\eta)$ acquires a
real part. Throughout this section
we work on the symmetric branch; the asymmetric family was treated in
Section~\ref{sec:asymmetric-quartic}.

Moreover,
\begin{equation}
 y(\ii\eta)
 =V'(\ii\eta)-2W(\ii\eta)
 \in\ii\mathbb{R}.
\end{equation}
Since $y^2=\Delta_{\mu,g}$, 
\begin{equation}
 \Delta_{\mu,g}(\ii\eta)\leq0.
 \label{eq:quartic-imaginary-discriminant-sign}
\end{equation}
Using $z^2=-q$, $z^4=q^2$, and $z^6=-q^3$, we find
\begin{equation}
 \Delta_{\mu,g}(\ii\eta)
 =-g^2q^3+2\mu gq^2-(1-4g)q-4\mu-4gm_2.
 \label{eq:quartic-discriminant-imaginary-axis}
\end{equation}
For $g>0$, every imaginary-axis node $z=\ii\eta$ therefore gives the lower bound
\begin{equation}
 m_2\geq L_{\mu,g}(q),
 \qquad
 L_{\mu,g}(q)
 =-\frac{\mu}{g}-\frac{g}{4}q^3
 +\frac{\mu}{2}q^2-\frac{1-4g}{4g}q.
 \label{eq:quartic-general-pick-envelope}
\end{equation}
A finite set of imaginary nodes gives
\begin{equation}
 m_2\geq\max_{1\leq j\leq L}L_{\mu,g}(\eta_j^2),
 \label{eq:quartic-finite-diagonal-nodes}
\end{equation}
whereas a continuously movable node gives the optimized envelope
\begin{equation}
 m_2\geq\mathcal{B}_{\mu}(g),
 \qquad
 \mathcal{B}_{\mu}(g)=\sup_{q>0}L_{\mu,g}(q).
 \label{eq:quartic-optimized-pick-bound}
\end{equation}
If the supremum lies at $q=0$, it is understood as the limit
$q\to0^+$, because the node itself must remain in the upper half-plane.

There is an important logical distinction.  The lower envelope
\eqref{eq:quartic-general-pick-envelope} follows from the reality of
$W(\ii\eta)$ on the symmetry axis, together with the quadratic loop
equation; it does not require the sign condition
$P(\ii\eta,\ii\eta)\geq0$.  Diagonal Pick positivity
instead fixes the sign of $\operatorname{Im}W(\ii\eta)$ and will
supply an upper constraint once the physical algebraic branch is selected by
continuation from infinity.  Thus reality and positivity are separate inputs:
$\Delta_{\mu,g}(\ii\eta)\leq0$ is necessary for a real symmetry-axis
solution, but it is not by itself sufficient to guarantee the Pick sign on
the physical sheet.  We therefore refer to \eqref{eq:quartic-optimized-pick-bound}
as the \emph{optimized symmetry-axis reality bound}.

\subsection{The stable quartic: \texorpdfstring{$\mu=+1$}{mu=+1}}

For the stable potential
\begin{equation}
 V(x)=\frac{x^2}{2}+\frac{g}{4}x^4,
 \qquad g>0,
\end{equation}
the imaginary-axis reality condition gives
\begin{equation}
 m_2\geq L_{+,g}(q),
 \qquad q=\eta^2,
\end{equation}
with
\begin{equation}
 L_{+,g}(q)
 =-\frac{1}{g}-\frac{g}{4}q^3
 +\frac{1}{2}q^2-\frac{1-4g}{4g}q.
 \label{eq:stable-quartic-envelope}
\end{equation}
Its derivative is
\begin{equation}
 L_{+,g}'(q)
 =-\frac{3g}{4}q^2+q-\frac{1-4g}{4g}.
 \label{eq:stable-quartic-envelope-derivative}
\end{equation}
Writing
\begin{equation}
 s=\sqrt{1+12g},
\end{equation}
the stationary points are
\begin{equation}
 q_{\pm}=\frac{2\pm s}{3g}.
\end{equation}
The point $q_-$, when positive, is a local minimum, while the global
maximum on $q\geq0$ is
\begin{equation}
 q_*=q_+=\frac{2+\sqrt{1+12g}}{3g}.
 \label{eq:stable-quartic-optimal-node}
\end{equation}
Reality therefore gives the lower bound
\begin{equation}
 m_2\geq\mathcal B_+(g)
 =\frac{(1+12g)^{3/2}-18g-1}{54g^2}.
 \label{eq:stable-quartic-sharp-lower-bound}
\end{equation}
By itself, this condition states only that $D(\ii\eta)\leq0$ along the
imaginary axis.  The physical Pick branch supplies
the complementary information.

Indeed, define the diagonal response
\begin{equation}
 P_\eta\equiv P(\ii\eta,\ii\eta)
 =-\frac{\operatorname{Im}W(\ii\eta)}{\eta}>0 .
\end{equation}
Since $W(\ii\eta)$ is purely imaginary on the symmetric branch,
\begin{equation}
 W(\ii\eta)=-\ii\eta P_\eta.
\end{equation}
For the stable quartic,
\begin{equation}
 V'(\ii\eta)=\ii\eta(1-g\eta^2),
\end{equation}
and hence
\begin{equation}
 y(\ii\eta)
 =\ii\eta\bigl[1-g\eta^2+2P_\eta\bigr].
 \label{eq:stable-quartic-y-imag-axis}
\end{equation}
For sufficiently small $\eta>0$, the quantity in brackets is positive:
$1-g\eta^2>0$ and $P_\eta>0$.  On the other hand, normalization of the
resolvent gives
\begin{equation}
 P_\eta
 =\frac{1}{\eta^2}-\frac{m_2}{\eta^4}+O(\eta^{-6}),
 \qquad \eta\to\infty,
\end{equation}
so that
\begin{equation}
 1-g\eta^2+2P_\eta<0
\end{equation}
for sufficiently large $\eta$.  The analytic function $y(z)$ is continuous
along the positive imaginary axis, and therefore there exists a
$\kappa>0$ for which
\begin{equation}
 y(\ii\kappa)=0.
 \label{eq:stable-quartic-forced-zero}
\end{equation}
Thus the nonbranching zero on the imaginary axis is not an additional
assumption: its existence is forced by the Pick sign near the real axis,
together with the physical normalization at infinity.

At this point
\begin{equation}
 D(\ii\kappa)=y(\ii\kappa)^2=0,
\end{equation}
so that
\begin{equation}
 m_2=L_{+,g}(\kappa^2)\leq \sup_{q>0}L_{+,g}(q)
 =\mathcal B_+(g).
\end{equation}
Combining this upper bound with
\eqref{eq:stable-quartic-sharp-lower-bound} gives
\begin{equation}
 m_2=\mathcal B_+(g)
 =\frac{(1+12g)^{3/2}-18g-1}{54g^2}.
 \label{eq:stable-quartic-exact-m2}
\end{equation}
In other words, imaginary-axis reality gives one side of the bound, while
global continuation of the physical Pick branch forces the curve to
touch the reality envelope and supplies the opposite side.  The second
moment is therefore fixed exactly.

Since $D=y^2$, the zero in \eqref{eq:stable-quartic-forced-zero} also obeys
\begin{equation}
 D'(\ii\kappa)=0.
\end{equation}
Equivalently, the touching point is stationary on the envelope.  Hence
\begin{equation}
 \kappa^2=q_*
 =\frac{2+\sqrt{1+12g}}{3g},
 \label{eq:stable-quartic-probe-double-zero}
\end{equation}
and the discriminant factorizes as
\begin{equation}
 D(z)
 =\bigl(gz^2+c_+\bigr)^2\bigl(z^2-B_+^2\bigr),
 \label{eq:stable-quartic-factorization}
\end{equation}
where
\begin{equation}
 c_+=\frac{\sqrt{1+12g}+2}{3},
 \qquad
 B_+^2=\frac{2\bigl(\sqrt{1+12g}-1\bigr)}{3g}.
\end{equation}
Thus the two nonbranching zeros are
\begin{equation}
 z=\pm\ii\kappa,
 \qquad
 \kappa^2=\frac{c_+}{g}=q_*.
\end{equation}
For generic $g>0$ the maximum of the envelope is nondegenerate, so these
are regular dressed saddles, i.e. double zeros of $D$.

The mechanism can be summarized geometrically.  If
$m_2<\mathcal B_+(g)$, then $D(\ii\eta)>0$ on part of the imaginary axis,
contradicting reality.  If $m_2>\mathcal B_+(g)$, then
$D(\ii\eta)<0$ strictly for all $\eta>0$, so $y(\ii\eta)$ cannot pass
through zero and its imaginary part cannot change continuously from the
Pick orientation near the real axis to the orientation required by
$y(z)\sim V'(z)-2/z$ at infinity.  The physical solution is therefore the
unique tangency
\begin{equation}
 m_2=\sup_{q>0}L_{+,g}(q),
 \qquad
 D(\ii\kappa)=D'(\ii\kappa)=0.
\end{equation}
In the Gaussian limit,
\begin{equation}
 q_*\sim\frac{1}{g}\to\infty,
 \qquad
 m_2=1-2g+O(g^2),
\end{equation}
which reduces continuously to the GUE value $m_2=1$.

\subsection{Negative quartic coupling and migration of the optimal probe}

We now consider the symmetric quartic model
\begin{equation}
V(x)=\frac{x^2}{2}+\frac{g}{4}x^4,
\qquad g<0.
\end{equation}
The real-axis integral is not convergent for $g<0$. Here we study its
formal one-cut continuation, whose compact density is nonnegative for
$-1/12\leq g<0$. Although the same algebraic expression for the second
moment continues to arise, the geometry of the relevant dressed saddles changes qualitatively.  For \(g>0\), the physical branch is forced to pass through a pair of imaginary dressed saddles.  For \(g<0\), this mechanism no longer operates on the imaginary axis; instead, the dressed saddles migrate to the real exterior region.

As before,
\begin{equation}
W(\ii\eta)=-\ii\eta P_\eta,
\qquad
P_\eta\equiv P(\ii\eta,\ii\eta)>0,
\end{equation}
and therefore
\begin{equation}
y(\ii\eta)
=
\ii\eta\left[1-g\eta^2+2P_\eta\right].
\end{equation}
For negative quartic coupling,
\begin{equation}
1-g\eta^2+2P_\eta
=
1+|g|\eta^2+2P_\eta>0
\end{equation}
for every \(\eta>0\).  Hence
\begin{equation}
\operatorname{Im}y(\ii\eta)>0,
\qquad
\eta>0,
\end{equation}
and there is no sign change along the positive imaginary axis.  In particular, analyticity does not force an imaginary dressed saddle in this case.

The imaginary-axis reality condition still gives a useful, but weak, constraint.  Writing
\begin{equation}
D(\ii\eta)
=
4g\left[L_{+,g}(\eta^2)-m_2\right],
\end{equation}
the condition
\begin{equation}
D(\ii\eta)\leq0
\end{equation}
now implies, because \(g<0\),
\begin{equation}
m_2\leq L_{+,g}(q),
\qquad q>0.
\end{equation}
For \(-1/12\leq g<0\), \(L_{+,g}(q)\) is increasing on \(q>0\), and therefore
\begin{equation}
m_2\leq -\frac{1}{g}.
\end{equation}
This bound alone does not determine the physical solution.

The relevant global branch-matching condition is instead found on the real axis outside the cut.  Let the physical support be
\begin{equation}
[-B,B].
\end{equation}
Immediately to the right of the endpoint \(B\), the physical branch satisfies
\begin{equation}
y(x)>0,
\qquad
B<x<B+\epsilon_0,
\end{equation}
for some $\epsilon_0>0$, whereas at large positive \(x\),
\begin{equation}
y(x)\sim V'(x)\sim gx^3<0.
\end{equation}
By continuity there is a point $r>B$ with $y(r)=0$.
Thus the physical branch is forced to contain a real exterior dressed saddle.

Since
\begin{equation}
D(r)=y(r)^2=0,
\end{equation}
we can express the second moment in terms of the real-axis envelope.  Writing
\begin{equation}
D(x)
=
4g\left[R_g(x^2)-m_2\right],
\qquad
R_g(t)=-\frac1g+\frac g4t^3+\frac12t^2
       +\frac{1-4g}{4g}t,
\label{eq:quartic-real-axis-envelope}
\end{equation}
reality in the exterior region gives
\begin{equation}
m_2\geq R_g(t),
\qquad
t=x^2>B^2,
\end{equation}
and hence
\begin{equation}
m_2\geq \sup_{t>B^2}R_g(t).
\end{equation}
On the other hand, the forced zero at \(x=r\) gives
\begin{equation}
m_2=R_g(r^2)
\leq
\sup_{t>B^2}R_g(t).
\end{equation}
Combining the two inequalities yields
\begin{equation}
m_2
=
\sup_{t>B^2}R_g(t).
\end{equation}
Thus, just as in the stable \(g>0\) case, global continuation of the physical branch forces saturation of the relevant reality envelope.

At the saturation point,
\begin{equation}
D(r)=0,
\qquad
D'(r)=0,
\end{equation}
so \(r\) is a regular dressed saddle.  For the quartic discriminant
\begin{equation}
D(z)
=
g^2z^6+2gz^4+(1-4g)z^2-4-4gm_2,
\end{equation}
the condition \(D'(r)=0\), with \(r\neq0\), gives
\begin{equation}
3g^2r^4+4gr^2+1-4g=0.
\end{equation}
Writing
\begin{equation}
q=r^2,
\end{equation}
the two solutions are
\begin{equation}
q_{\pm}
=
\frac{-2\pm\sqrt{1+12g}}{3g}.
\end{equation}
For
\begin{equation}
-\frac{1}{12}<g<0,
\end{equation}
the physical exterior saddle is the larger root,
\begin{equation}
r^2
=
\frac{-2-\sqrt{1+12g}}{3g}.
\end{equation}
Substituting this into \(D(r)=0\) gives
\begin{equation}
m_2
=
\frac{(1+12g)^{3/2}-18g-1}{54g^2}.
\end{equation}
This is the same algebraic expression obtained for \(g>0\), but the corresponding dressed-saddle geometry is different.

The two regimes are connected by migration through infinity:
\begin{equation}
\begin{aligned}
g>0 &: \quad z=\pm\ii\kappa,\\
g\to0 &: \quad |z|\to\infty,\\
-\frac1{12}<g<0 &: \quad z=\pm r\in\mathbb R.
\end{aligned}
\end{equation}
At $g=-1/12$, the exterior dressed saddles collide with the branch points.  The discriminant becomes
\begin{equation}
D(z)
=
\frac{1}{144}(z^2-8)^3,
\end{equation}
so the double zeros cease to be nonbranching and instead become cubic critical branch points.  The negative-coupling endpoint therefore marks the breakdown of the regular dressed-saddle geometry and the onset of a critical degeneration of the cut structure.

\subsection{The double-well quartic: \texorpdfstring{$\mu=-1$}{mu=-1}}

Consider
\begin{equation}
 V(x)=-\frac{x^2}{2}+\frac{g}{4}x^4,
 \qquad g>0.
\end{equation}
For $0<g<1/4$ the physical symmetric solution has two cuts,
\begin{equation}
 [-B,-A]\cup[A,B].
\end{equation}
Oddness of the resolvent and analyticity in the central gap imply
\begin{equation}
 W(0)=0.
\end{equation}
Since
\begin{equation}
 \mathcal Q(z)=-1+g(z^2+m_2),
\end{equation}
the loop equation at the origin gives
\begin{equation}
 m_2=\frac1g,\qquad 0<g<\frac14.
 \label{eq:m2twocut}
\end{equation}
The corresponding discriminant factorizes as
\begin{equation}
 D(z)=g^2z^2(z^2-A^2)(z^2-B^2),
\end{equation}
where
\begin{equation}
 A^2=\frac{1-2\sqrt g}{g},
 \qquad
 B^2=\frac{1+2\sqrt g}{g}.
\end{equation}
Thus the origin is a regular dressed saddle in the central gap.

The same value has a useful finite-plane interpretation.  On the imaginary
axis, $y(\ii\eta)$ is purely imaginary and hence $D(\ii\eta)\leq0$.  With
$q=\eta^2$ this gives
\begin{equation}
 m_2\geq L_{-,g}(q),
 \qquad
 L_{-,g}(q)
 =\frac1g-\frac g4q^3-\frac12q^2
 -\frac{1-4g}{4g}q.
 \label{eq:double-well-quartic-envelope}
\end{equation}
For $0<g<1/4$, $L_{-,g}(q)$ decreases for $q>0$, so
\begin{equation}
 m_2\geq\sup_{q>0}L_{-,g}(q)=\frac1g.
\end{equation}
On the real central gap, $D(x)=y(x)^2\geq0$, and evaluating at $x=0$
gives the opposite inequality
\begin{equation}
 m_2\leq\frac1g.
 \label{eq:double-well-real-gap-upper}
\end{equation}
The exact two-cut value is therefore the meeting point of the two reality
conditions.  In this symmetric example the equality is already fixed by
$W(0)=0$; the two-sided reality pinch is its geometric manifestation.

At $g=1/4$ the inner endpoints reach the origin.  The dressed saddle and
the two inner branch points then coalesce, and the discriminant develops a
fourth-order zero,
\begin{equation}
 D(z)\sim z^4.
\end{equation}
For $g>1/4$ the support becomes one cut and the real central-gap condition
is no longer available.  The imaginary-axis envelope instead develops an
interior maximum at
\begin{equation}
 q_*=\frac{\sqrt{1+12g}-2}{3g},
 \label{eq:double-well-one-cut-optimal-node}
\end{equation}
and reality gives
\begin{equation}
 m_2\geq\mathcal B_-(g)
 =\frac{(1+12g)^{3/2}+18g+1}{54g^2}.
 \label{eq:double-well-one-cut-reality-bound}
\end{equation}
We now show that the physical Pick branch forces this bound to be
saturated.

\subsection{Diagonal Pick positivity and branch selection}

On the symmetry axis define
\begin{equation}
 P_\eta\equiv P(\ii\eta,\ii\eta)
 =-\frac{\operatorname{Im}W(\ii\eta)}{\eta}>0.
\end{equation}
Since $W(\ii\eta)=-\ii\eta P_\eta$, the spectral function is
\begin{equation}
 y(\ii\eta)
 =\ii\eta\bigl[-1-g\eta^2+2P_\eta\bigr].
 \label{eq:double-well-y-imag-axis}
\end{equation}
This formula fixes the orientation of the physical branch along the
imaginary axis.

In the two-cut phase the physical factorization can be written as
\begin{equation}
 y(z)=gz\sqrt{(z^2-A^2)(z^2-B^2)},
\end{equation}
where the square root is chosen to behave as $z^2$ at infinity.  It follows
that
\begin{equation}
 y(\ii\eta)
 =-\ii g\eta\sqrt{(\eta^2+A^2)(\eta^2+B^2)},
 \qquad \eta>0,
\end{equation}
so $\operatorname{Im}y(\ii\eta)<0$ throughout the positive imaginary axis.
The only dressed saddle relevant to this direction is therefore the one at
the boundary point $z=0$.  On the real central gap the same analytic branch
changes sign across the origin; locally,
\begin{equation}
 y(z)=-\sqrt{1-4g}\,z+O(z^3).
\end{equation}
This is consistent with the two-sided reality pinch described above.

The one-cut phase is different.  The origin now lies inside the support, so
\begin{equation}
 \eta P_\eta=-\operatorname{Im}W(\ii\eta)
 \longrightarrow \pi\rho(0)>0,
 \qquad \eta\to0^+.
\end{equation}
Equation~\eqref{eq:double-well-y-imag-axis} therefore gives
\begin{equation}
 y(\ii0^+)=2\pi\ii\rho(0),
\end{equation}
so the imaginary part of $y$ is positive close to the real axis.  On the
other hand,
\begin{equation}
 y(\ii\eta)\sim V'(\ii\eta)=-\ii g\eta^3,
 \qquad \eta\to\infty,
\end{equation}
and its imaginary part is negative at large $\eta$.  Continuity of the
physical analytic branch therefore forces a zero
\begin{equation}
 y(\ii\kappa)=0
 \label{eq:double-well-forced-imag-zero}
\end{equation}
for some $\kappa>0$.

Since $D=y^2$, this gives
\begin{equation}
 D(\ii\kappa)=0
 \quad\Longrightarrow\quad
 m_2=L_{-,g}(\kappa^2).
\end{equation}
But $L_{-,g}(\kappa^2)$ cannot exceed its maximum, and hence
\begin{equation}
 m_2\leq\sup_{q>0}L_{-,g}(q)=\mathcal B_-(g).
\end{equation}
Together with the reality bound
\eqref{eq:double-well-one-cut-reality-bound}, this fixes
\begin{equation}
 m_2=\mathcal B_-(g)
 =\frac{(1+12g)^{3/2}+18g+1}{54g^2},
 \qquad g>\frac14.
 \label{eq:double-well-one-cut-exact}
\end{equation}
Thus the optimized reality envelope is saturated because the physical
Pick branch must reverse its orientation between the real axis and
infinity.

Since the equality is attained at the maximum of $L_{-,g}$,
\begin{equation}
 \kappa^2=q_*
 =\frac{\sqrt{1+12g}-2}{3g},
\end{equation}
and
\begin{equation}
 D(\ii\kappa)=D'(\ii\kappa)=0.
\end{equation}
For generic $g>1/4$ this is a regular dressed saddle.  The one-cut
discriminant takes the form
\begin{equation}
 D(z)=\bigl(gz^2+c_-\bigr)^2(z^2-B_-^2),
\end{equation}
with
\begin{equation}
 c_-=\frac{\sqrt{1+12g}-2}{3},
 \qquad
 B_-^2=\frac{2(\sqrt{1+12g}+1)}{3g}.
\end{equation}
The two dressed saddles are therefore $z=\pm\ii\kappa$.

\subsection{Analyticity as an algebraic constraint}

The branch argument above has a simple algebraic interpretation.  A physical
Hermitian resolvent is analytic in the upper half-plane, and therefore
$y(z)=V'(z)-2W(z)$ is analytic there as well. Consequently every zero
of $D(z)=y(z)^2$ in the upper half-plane must have even multiplicity.  An odd-order zero of
$D$ would instead be a branch point of $W$ and is inadmissible away from the
real spectral support.

For the one-cut double-well quartic this criterion is already encoded in the
reality envelope.  If
\begin{equation}
 m_2<\mathcal B_-(g),
\end{equation}
then $D(\ii\eta)>0$ on part of the imaginary axis, contradicting the fact that
$y(\ii\eta)$ is purely imaginary.  If
\begin{equation}
 m_2>\mathcal B_-(g),
\end{equation}
then $D(\ii\eta)<0$ strictly for every $\eta>0$, so $y(\ii\eta)$ cannot vanish.
This is incompatible with the change of sign required by the physical
Pick branch.  The unique admissible value is therefore the tangency
\begin{equation}
 D(\ii\kappa)=D'(\ii\kappa)=0,
\end{equation}
which is the same condition as saturation of the imaginary-axis envelope.

The transition at $g=1/4$ can now be described entirely in terms of the
zeros of the spectral curve.  For $g<1/4$ the origin is a regular dressed
saddle lying in the real central gap.  At $g=1/4$ it collides with the two
inner branch points, producing a fourth-order zero of $D$.  For $g>1/4$
this critical zero resolves into the conjugate pair of regular nonbranching
zeros $z=\pm\ii\kappa$.  Thus the optimal Cauchy scale follows the same
spectral reorganization that accompanies the closing of the gap.

\subsection{Relation to the moment bootstrap}
\label{sec:quartic-diagonal-pick-versus-hankel}

Section~\ref{subsec:information-matched} showed that, at matched finite
polynomial information, truncated Cauchy probes and the Hankel basis
define the same positivity cone.  The quartic model provides a simple
illustration of how the two organizations nevertheless expose different
features of the exact solution.

On the symmetric branch, positivity of the polynomial basis
$\{1,x^2\}$ gives
\begin{equation}
 \begin{pmatrix}
 1&m_2\\
 m_2&m_4
 \end{pmatrix}\succeq0.
\end{equation}
Using the loop equation
\begin{equation}
 \mu m_2+gm_4=1,
\end{equation}
one obtains
\begin{equation}
 gm_2^2+\mu m_2-1\leq0,
\end{equation}
and hence, for $g>0$,
\begin{equation}
 m_2\leq\frac{\sqrt{1+4g}-\mu}{2g}.
 \label{eq:quartic-low-hankel-upper-bound}
\end{equation}
Higher Hankel matrices refine this bound in the usual nested hierarchy.

The exact Pick response contains the same moment data but reorganizes
it at finite $z$.  In the quartic examples above, the distinguished
probe location is the dressed saddle selected by the analytic structure
of the physical branch.  The point of the rational formulation is
therefore not a stronger finite positivity principle, but a direct
finite-plane localization of the spectral feature at which the
constraint is saturated.

\bibliographystyle{JHEP}
\bibliography{refs}

\end{document}